\documentclass[preprint,superscriptaddress,amsmath,amssymb,aps,floatfix,nofootinbib,longbibliography]{revtex4-1}

\usepackage[pdftex]{graphicx}
\usepackage{bm}
\usepackage{color}
\usepackage{times}
\usepackage{amsmath}
\usepackage{amssymb}
\usepackage{mathtools}
\newcommand{\defeq}{\vcentcolon=}

\begin{document}

\title{From arteries to boreholes: Transient response \\ of a poroelastic cylinder to fluid injection}

\author{Lucy C. Auton}
\affiliation{Department of Engineering Science, University of Oxford, Oxford, OX1 3PJ, UK}
\author{Christopher W. MacMinn}
\email{christopher.macminn@eng.ox.ac.uk}
\affiliation{Department of Engineering Science, University of Oxford, Oxford, OX1 3PJ, UK}

\date{\today}


\begin{abstract}
The radially outward flow of fluid through a porous medium occurs in many practical problems, from transport across vascular walls to the pressurisation of boreholes in the subsurface. When the driving pressure is non-negligible relative to the stiffness of the solid structure, the poromechanical coupling between the fluid and the solid can control both the steady-state and the transient mechanics of the system. Very large pressures or very soft materials lead to large deformations of the solid skeleton, which introduce kinematic and constitutive nonlinearity that can have a nontrivial impact on these mechanics.
Here, we study the transient response of a poroelastic cylinder to sudden fluid injection. We consider the impacts of kinematic and constitutive nonlinearity, both separately and in combination, and we highlight the central role of driving method in the evolution of the response. We show that the various facets of nonlinearity may either accelerate or decelerate the transient response relative to linear poroelasticity, depending on the boundary conditions and the initial geometry, and that an imposed fluid pressure leads to a much faster response than an imposed fluid flux.
\end{abstract}

\maketitle

\section{Introduction}

Radial flow of fluid through a porous material plays a key role in many practical problems in, for example, geomechanics, biophysics, and filtration. Scenarios involving large injection pressures, soft materials, or thin structures may result in large deformations that introduce both kinematic and constitutive nonlinearity. This nonlinearity can have nontrivial impacts on both the steady-state and the transient mechanics of the system. Here, we consider these impacts in the context of a model problem: The response of a poroelastic cylinder to sudden, radially outward fluid injection. We previously derived the general nonlinear model for this problem under the assumption of incompressible constituents, and we used this model to study the deformation at steady state~\citep{auton2017arteries}. Specifically, we investigated the impact of geometry, permeability law, and outer boundary condition for classical linear poroelasticity, fully nonlinear poroelasticity, and an intermediate model. We now use the same model to study the transient evolution of the deformation. The transient evolution is particularly important in biomedical applications because these systems are inherently transient, as with the periodic pressure pulses in vascular flows, and in geophysical applications where the primary interest is in the time needed for a certain amount of consolidation or pressure dissipation to occur.

The transient aspects of large-deformation poroelasticity have been considered in some detail in a rectilinear (uni-axial) geometry in the context of fluid injection~\citep{macminn2016large, barry1991unsteady}, forced infiltration~\cite{sommer1996forced, preziosi1996infiltration}, and transmural flow~\cite{kenyon1979mathematical}. For example, \citet{macminn2016large} considered the effect of constant versus deformation-dependent permeability in a model problem where the deformation is strictly compressive, showing that deformation-dependent permeability can greatly increase the evolution timescale (slow the response) relative to constant permeability.

The radial geometry has attracted interest in a range of contexts, including biomedical applications such as subcutaneous injections and flow through arterial walls~\citep[\textit{e.g},][]{van2012needle, jayaraman1983water} and geophysical applications such as borehole pressurisation or consolidation following pile driving~\citep[\textit{e.g},][]{randolph1979analytical, rice1976some}. For scenarios involving small deformations, it is appropriate to use classical linear poroelasticity; this leads to a linear partial differential equation (PDE) that can be solved analytically via various classical methods. \citet{kenyon1976transient} and \citet{jayaraman1983water} considered fluid flow across arterial walls, in which the artery is modelled as a soft porous cylinder with a time-dependent inner fluid pressure and a constrained outer boundary. \citet{kenyon1976transient} considered the transient response to a step change in the inner pressure and used a Laplace transform to derive an approximate solution for small times. \citet{jayaraman1983water} considered the transient response to an oscillatory inner fluid pressure, using normal modes to derive approximate long-time solutions for low- and high-frequency driving. \citet{randolph1979analytical} considered consolidation after pile-driving, in which the insertion of the pile leads to locally elevated pore pressure that subsequently dissipates. They modelled the relaxation of the soil around the cylindrical pile, treating the pile as a rigid and impermeable boundary from which the soil could not separate and assuming that pore pressure was only perturbed from its initial value in a finite region around the pile. They used separation of variables to derive a solution in terms of an infinite series of Bessel functions. \citet{jana1963deformation} considered elastic deformations about a cylindrical cavity in an infinite medium, deriving solutions via Fourier Series and Laplace Transforms. Similarly, \citet{detournay1988poroelastic} considered relaxation around an excavated or pressurised borehole by modelling the borehole as a cylindrical cavity in an infinite domain, with various modes of loading at the interface. They applied a Laplace transform in time, solved the spatial problem in terms of modified Bessel functions, and then inverted the Laplace transform numerically. \citet{rice1976some} considered the internal pressurisation of annular rock specimens, modelling these as unconstrained cylinders in plane strain. They solved this problem using complex variables, deriving an approximate solution for small times and a complete solution for the limiting case of a semi-infinite domain.

The introduction of nonlinearity leads to a problem that is less analytically tractable. \citet{barry1993radial} and \citet{barry1998effect} accounted partially for large deformations by including moving boundaries and deformation-dependent permeability in a model that was otherwise linearised. \citet{barry1993radial} considered a similar model for a constrained cylinder, using perturbation methods to derive approximate solutions for small times and for slow compression rates. \citet{barry1998effect} again used a similar model to study constrained and unconstrained cylinders and develop approximate solutions for small times.

These previous works have considered a wide range of applications and model problems, but a systematic exploration of the transient mechanics of poroelastic cylinders is still lacking. Here, we consider axisymmetric deformations due to sudden fluid injection into the inner cavity of a poroelastic cylinder in plane strain, again assuming incompressible constituents \citep{auton2017arteries}. We define the general parameter space and explore the effects of geometry, outer boundary condition, and driving method (fixed pressure difference or fixed flow rate), as well as the impact of large deformations, on the transient response. For classical linear poroelasticity, we use separation of variables and Sturm-Liouville theory to derive analytical solutions for the fixed-flow–rate problem in terms of an infinite series of Bessel and Struve functions. We solve the various nonlinear problems numerically using the method of lines with Chebyshev spectral collocation \citep{piche2007solving}, a natural extension of the pseudospectral method used in \citet{auton2017arteries}. We examine the transient evolution of the deformation and show that the evolution timescale exhibits a complex dependence on geometry, boundary conditions, and driving method, even for linear poroelasticity. For the range of cases considered here, nonlinear elasticity has a much smaller impact on the evolution timescale than nonlinear kinematics, deformation-dependent permeability, or the other factors mentioned previously.

\section{Model problem}

\begin{figure}[tb]
    \centering
    \includegraphics[width=5in]{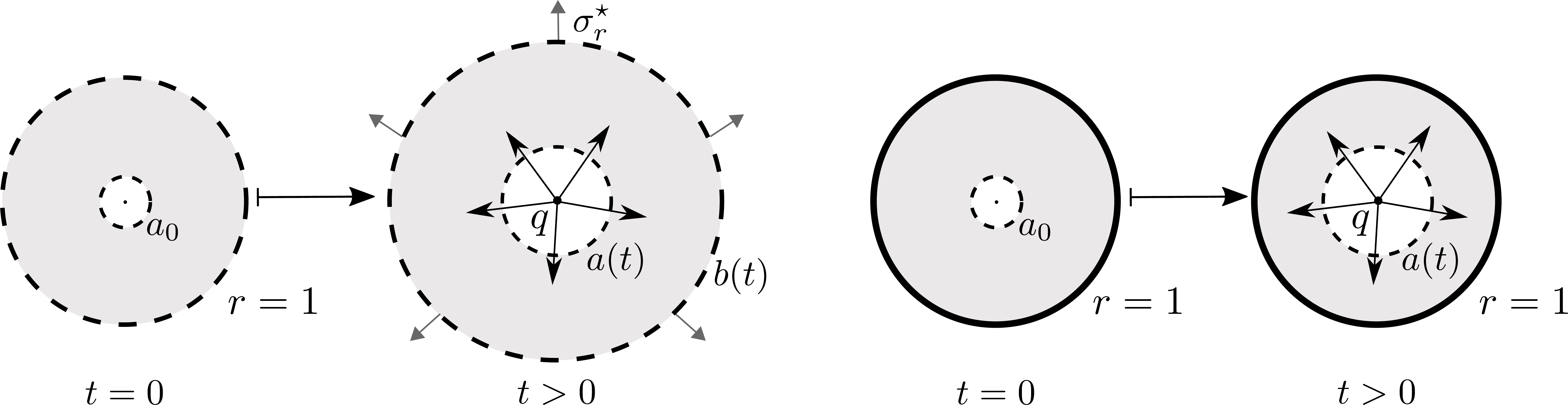}
    \caption{We consider the transient response of a poroelastic cylinder to sudden fluid injection. The cylinder has dimensionless initial inner radius $r=a_0$ and dimensionless initial outer radius $r=1$. The inner boundary is free to move, while the outer boundary is either subject to a fixed radial effective stress $\sigma_r^\star$ (left) or fixed in place (right). We initialise the problem from a relaxed state and we study the transient evolution of the problem towards its steady state. Note that we assume plane strain and adopt the convention of tension being positive. \label{fig:rad} }
\end{figure}

We consider radially outward fluid injection into a poroelastic cylinder from a line source, which is characterised by either a fixed pressure difference or a fixed flow rate. We consider two distinct outer boundary conditions on the cylinder: Fixed radial effective stress (Figure~\ref{fig:rad}, left) and zero displacement (Figure~\ref{fig:rad}, right).

\subsection{Summary of Theory}

We now summarise the model in dimensionless form, denoting dimensional quantities with a tilde. For the full derivation of the axisymmetric problem in dimensional form, see \citet{auton2017arteries}. For a full three-dimensional presentation, see \citet{macminn2016large}.

\subsubsection{Scaling}\label{scaling}

To write the model in dimensionless form, we adopt characteristic scales for length, stress/pressure, time, and permeability. We take the dimensional initial outer radius $\tilde{b}_0$ to be the characteristic length scale and the $p$-wave (oedometric) modulus $\tilde{\mathcal{M}}$ to be the characteristic stress/pressure scale, and we adopt the classical poroelastic timescale $\tilde{T}_\mathrm{pe}\defeq \tilde{b}_0^2\tilde{\mu}/\tilde{k}_0\tilde{\mathcal{M}}$, where $\tilde{\mu}$ is the dynamic viscosity of fluid and $\tilde{k}_0$ is the characteristic permeability scale. We model fluid injection as a line source at the origin, characterised by either a fixed dimensionless flow rate $q$ or a fixed dimensionless pressure difference $\Delta{p}\defeq{}p(a,t)-p(b,t)$, where $p(r,t)$ is the fluid (pore) pressure at radial position $r$ and time $t$, and $a=a(t)$ and $b=b(t)$ are the inner and outer radii of the cylinder, respectively.

The dimensionless model is then characterised by the reference (relaxed) porosity $\phi_{f,0}$, which we take to be uniform for simplicity, and by four other dimensionless parameters:
\begin{equation}
    \Gamma\defeq\frac{\tilde{\Lambda}}{\tilde{\mathcal{M}}}, \quad
    a_0 \defeq\frac{\tilde{a}_0}{{\tilde{b}_0}}, \quad
    \sigma_r^\star\defeq\frac{\tilde{\sigma}_r^\star}{\tilde{\mathcal{M}}}, \quad
    \text{and either} \quad
    q\defeq\frac{\tilde{\mu}\tilde{Q}}{2\pi\tilde{k}_0\tilde{\mathcal{M}}} \quad
    \text{or} \quad
    \Delta{p}\defeq\frac{\Delta\tilde{p}}{\tilde{\mathcal{M}}},
\end{equation}
where $\tilde{\Lambda}$ is Lam\'{e}'s first parameter, $a_0$ is the dimensionless initial inner radius, ${\sigma}_r^\star$ is the dimensionless radial effective stress at the outer boundary, and $\tilde{Q}(t)$ is the dimensional volume injection rate per unit length into the page. Note that only one of $q$ or $\Delta{p}$ can be imposed --- the other evolves in time as part of the solution (\textit{c.f.,} \S\ref{driving}).

\subsubsection{Kinematics}

For axisymmetric flow and deformation, the fluid velocity $\mathbf{v}_f$, the solid displacement $\mathbf{u}_s$, and the solid velocity $\mathbf{v}_s$ are strictly in the radial direction, and are functions of only the Eulerian radial co-ordinate $r$ and time $t$, 
\begin{equation}
    \mathbf{v}_f=v_f(r,t)\hat{\mathbf{e}}_r, \quad
    \mathbf{u}_s=u_s(r,t)\hat{\mathbf{e}}_r, \quad
    \text{and} \quad
    \mathbf{v}_s=v_s(r,t)\hat{\mathbf{e}}_r,
\end{equation}
where $\hat{\mathbf{e}}_r$ is the radial unit vector. The deformation is characterised by the three principal stretch ratios $\lambda_r$, $\lambda_\theta$, and $\lambda_z$, which for axisymmetry and plane strain are given by 
\begin{equation}\label{stretches}
    \lambda_r=\left(1-\frac{\partial u_{s}}{\partial{r}}\right)^{-1}, \quad
    \lambda_\theta=\left(1-\frac{u_{s}}{r}\right)^{-1}, \quad
    \text{and} \quad
    \lambda_z\equiv{}1.
\end{equation}
The Jacobian determinant $J$ measures the local volume change,
\begin{equation}
    J(r,t)=\lambda_r\lambda_\theta\lambda_z=\lambda_r\lambda_\theta.
\end{equation}
Hence, under the assumption that the solid and fluid phases are individually incompressible, deformation must occur through rearrangement of the solid skeleton with corresponding changes in the local porosity, $\phi_f$, giving 
\begin{equation}\label{J}
    J(r,t)=\frac{1-\phi_{f,0}}{1-\phi_f}.
\end{equation}
Equations~(\ref{stretches})--(\ref{J}) lead to a kinematic relationship between porosity and displacement,
\begin{equation}\label{eq:phi_to_u}
    \frac{\phi_f-\phi_{f,0}}{1-\phi_{f,0}} =\frac{1}{r}\frac{\partial}{\partial{r}}\left(ru_s-\frac{1}{2}u_s^2\right).
\end{equation}
Finally, local conservation of mass for the fluid and solid constituents is given by
\begin{equation}\label{cons}
    \frac{\partial{\phi_f}}{\partial{t}}+\frac{1}{r}\frac{\partial}{\partial{r}} \left(r\phi_fv_{f}\right)=0 \quad
    \mathrm{and} \quad
    \frac{\partial{\phi_f}}{\partial{t}}-\frac{1}{r}\frac{\partial}{\partial{r}} \big[r(1-\phi_f)v_{s}\big]=0,
\end{equation}
respectively, and the fluid and solid velocities are related to the injection rate via
\begin{equation}
    \phi_fv_f+(1-\phi_f)v_s = \frac{q}{r}.
\end{equation}

\subsubsection{Mechanical equilibrium}

In the absence of body forces and neglecting inertia, mechanical equilibrium is given by
\begin{equation}\label{Cauchy}
    \frac{\partial{\sigma^\prime_r}}{\partial{r}} +\frac{\sigma^\prime_r-\sigma^\prime_{\theta}}{r}=\frac{\partial{p}}{\partial{r}},
\end{equation}
where $\sigma^\prime_r$ and $\sigma^\prime_{\theta}$ are the radial and azimuthal components, respectively, of Terzaghi's effective Cauchy stress (\textit{i.e.}, the stress supported by the solid through deformation). Note that we take tension to be positive.

\subsubsection{Darcy's Law}

We assume that the fluid flows relative to the solid skeleton according to Darcy's law. In the absence of body forces, this is written
\begin{equation}\label{Darcy}
    \phi_f(v_f-v_s)=-k(\phi_f)\frac{\partial{p}}{\partial{r}},
\end{equation}
where $k(\phi_f)\equiv\tilde{k}(\phi_f)/\tilde{k}_0$ is the dimensionless permeability, which we take to be an isotropic function of porosity (see \S\ref{perm}). Equations~(\ref{cons}) and (\ref{Darcy}) lead to a conservation law in $\phi_f$,
\begin{equation}\label{govern}
    \frac{\partial \phi_f}{\partial{t}}+\frac{1}{r}\frac{\partial}{\partial{r}}\left[q\phi_f-r(1-\phi_f)k(\phi_f)\frac{\partial p}{\partial{r}}\right] = 0, 
\end{equation}
and to two expressions for the solid velocity, 
\begin{equation}
\label{Vel}
v_s = \frac{q}{r}+k(\phi_f)\frac{\partial p}{\partial{r}} \quad \text{and}\quad v_s = \lambda_r \frac{\partial{u_s}}{\partial{t}}.
\end{equation}
The former combines Darcy's Law with conservation of mass; the later is strictly kinematic.

\subsubsection{Elasticity laws}

As in \citet{auton2017arteries}, we consider two elastic constitutive relations: Hencky elasticity and linear elasticity. Hencky elasticity is a generic constitutive law that captures the kinematic aspects of large deformations without introducing material-specific complexity. Hencky elasticity also uses the same two elastic parameters as linear elasticity, and is asymptotically equivalent to linear elasticity in the limit of infinitesimal strain. Hencky elasticity is therefore a convenient and appropriate model for a wide range of materials under moderate deformations \citep{hencky1931law, anand1979h, xiao2002hencky, bazant1998easy}. Note that, whereas moderate to large elastic deformations are common in the context of soft filters and tissues, soils and rocks are unlikely to behave elastically beyond small strains. Hencky elasticity will be appropriate for soils and rocks up to the point of brittle or ductile failure.

For Hencky elasticity, the radial and azimuthal components of the effective Cauchy stress are
\begin{equation}\label{henckconst}
    \sigma_r^\prime=\frac{\ln{\lambda_r}}{J} +\Gamma\frac{\ln{\lambda_\theta}}{J} \quad
    \text{and} \quad
    \sigma_\theta^\prime=\Gamma\frac{\ln{\lambda_r}}{J} +\frac{\ln{\lambda_\theta}}{J}.
\end{equation}
Combining the above with Equation~(\ref{Cauchy}) we obtain
\begin{equation}\label{dpdrhenck}
    \frac{\partial{p}}{\partial{r}}=\frac{\partial}{\partial{r}}\left[\frac{\ln\left( \lambda_r\lambda_\theta^\Gamma\right)}{J}\right]+\frac{1-\Gamma}{Jr}\left[\ln\left(\frac{\lambda_r}{\lambda_\theta}\right)\right].
\end{equation}
For linear elasticity, the relevant components of effective Cauchy stress are instead given by
\begin{equation}\label{eq:sigma_small_rad}
    \sigma_r^\prime=\frac{\partial{u_s}}{\partial{r}} +\Gamma\displaystyle\frac{u_s}{r} \quad
    \text{and} \quad
    \sigma_\theta^\prime=\Gamma\frac{\partial{u_s}}{\partial{r}} +\displaystyle\frac{u_s}{r}.
\end{equation}
Combining this with Equation~(\ref{Cauchy}) leads to
\begin{equation}\label{dpdrlin}
    \frac{\partial{p}}{\partial{r}} =\frac{\partial}{\partial{r}}\left[\frac{1}{r}\frac{\partial}{\partial{r}}(u_sr)\right].
\end{equation}

\subsubsection{Permeability Laws}\label{perm}

Deformation of the solid skeleton will alter the pore structure and is thus likely to change the permeability. To capture this effect, we adopt a normalised Kozeny-Carman permeability law,
\begin{equation}\label{KC}
    k(\phi_f)=\frac{(1-\phi_{f,0})^2}{\phi_{f,0}^3} \left[\frac{\phi_f^3}{(1-\phi_f)^2}\right],
\end{equation}
such that $k(\phi_{f,0})=1$. This expression encapsulates the qualitatively important properties that the permeability vanishes as $\phi_f$ tends to zero and diverges as $\phi_f$ tends to one. This nonlinear effect is neglected in classical linear poroelasticity, in which case the permeability function is simply $k(\phi_f)=k(\phi_{f,0})\equiv{}1$. Below, we denote models using deformation-dependent permeability (Equation~(\ref{KC})) by `-$k_\mathrm{KC}$' and those using constant permeability by `-$k_0$'.

\subsubsection{Linearisation and model summary}

As in \citet{auton2017arteries}, we adopt three classes of models. Fully nonlinear models combine exact kinematics (Equations~(\ref{eq:phi_to_u}), (\ref{govern}), and (\ref{Vel})) with Hencky elasticity (Equations~(\ref{henckconst})--(\ref{dpdrhenck})). Intermediate models combine exact kinematics (Equations~(\ref{eq:phi_to_u}), (\ref{govern}), and (\ref{Vel})) with linear elasticity (Equation~\ref{eq:sigma_small_rad}). We refer to the fully nonlinear class of models as `N models' and to the intermediate class of models as `Q models'. The third class of models is based on linear poroelasticity.

Classical linear poroelasticity relies on the assumption of infinitesimal deformations, which, here, corresponds to the assumption of infinitesimal strains, $u_s/r\ll{}1$ and $\partial{u_s}/\partial{r}\ll{}1$. The kinematics can be simplified under this assumption such that Equations (\ref{eq:phi_to_u}), (\ref{govern}), and (\ref{Vel}) become
\begin{equation}\label{eq:phi_to_ulin}
    \frac{\phi_f-\phi_{f,0}}{1-\phi_{f,0}} \approx\frac{1}{r}\frac{\partial}{\partial{r}}\left(ru_s\right), \quad
    \frac{\partial{\phi_f}}{\partial{t}}-\frac{1}{r}\frac{\partial}{\partial{r}} \left(rk(\phi_f)(1-\phi_{f,0})\frac{\partial{p}}{\partial{r}}\right)\approx0, \quad
    \mathrm{and} \quad
    v_s\approx\frac{\partial{u_s}}{\partial{t}},
\end{equation}
respectively. We refer to the class of models combining Equations~(\ref{eq:phi_to_ulin}) with Equations~(\ref{eq:sigma_small_rad}) and (\ref{dpdrlin}) as `L models', such that the L-$k_0$ model is classical linear poroelasticity. We also consider an L-$k_\mathrm{KC}$ model that, although nonlinear in the strain and therefore asymptotically inconsistent, allows us to isolate the impact of deformation-dependent permeability from those of nonlinear kinematics and nonlinear elasticity.

\subsection{Initial, boundary and driving conditions}\label{ICandBCS}

\subsubsection{Boundary Conditions}

The inner boundary is free to move, subject to no radial effective stress,
\begin{equation}\label{innerBC}
    \sigma^\prime_r(a,t)=0.
\end{equation}
The inner boundary is also a material boundary, subject to the kinematic conditions\footnote{Note that \citet{auton2017arteries} erroneously state that $v_s(a,t)=\mathrm{d}a/\mathrm{d}t= \frac{\partial{u_s}}{\partial{t}}\Big|_{r=a}$. The latter equality is incorrect.}
\begin{equation}\label{innerKBC}
    u_{s}(a,t)=a(t)-a_0 \quad
    \text{and} \quad
    v_s(a,t)=\frac{\mathrm{d}a}{\mathrm{d}t}.
\end{equation}
At the outer boundary, we enforce a vanishing pressure without loss of generality
\begin{equation}\label{p}
    p(b,t)=0.
\end{equation}
Additionally, we consider the same two distinct sets of outer boundary conditions as in~\citet{auton2017arteries}, imposing either a fixed position (no displacement) or a fixed radial effective stress,
\begin{equation}\label{outerBCb}
    u_s(b,t)=0 \quad
    \mathrm{or} \quad
    \sigma_r^\prime(b,t)=\sigma_r^\star.
\end{equation}
We refer to the former as `constrained' and to the latter, for the limiting case of $\sigma_r^\star\equiv{}0$, as `unconstrained'. The latter case involves a moving boundary, and is therefore additionally subject to the kinematic conditions
\begin{equation}\label{outerKBC}
    u_{s}(b,t)=b(t)-b_0 \quad
    \text{and} \quad
    v_s(b,t)=\frac{\mathrm{d}b}{\mathrm{d}t}.
\end{equation}
Note that, for the L models, the inner and outer conditions are applied at $a=a_0$ and $b=1$, respectively.

For convenience, we define three operators
\begin{equation}\label{bclucy}
    B^{a}[u_s]\defeq\sigma^\prime_r(a,t), \quad
    B_1^b[u_s]\defeq{}u_s(b,t), \quad
    \text{and} \quad
    B^{b}_2[u_s]\defeq\sigma'_r(b,t)-\sigma^\star_r.
\end{equation}
So that Equation~(\ref{innerBC}) can be expressed as $B^{a}[u_s]=0$ and Equation~(\ref{outerBCb}) can be expressed as $B_i^b[u_s]=0$ for $i=1,2$, where $i=1$ corresponds to a constrained cylinder and $i=2$ corresponds to a cylinder subject to a fixed radial effective stress at the outer boundary. In the limiting case of $\sigma^\star_r\equiv{}0$, $i=2$ corresponds to an unconstrained cylinder.

\subsubsection{Initial conditions}

We initialise the model by specifying $u_s(r,0)=u_{s,0}(r)$ and assuming that the cylinder starts from rest $(v_s(r,0)=v_f(r,0)=0)$. For both the constrained and unconstrained cylinders, we start from a relaxed (undeformed and stress-free) state, such that $u_{s,0}=0$. This implies that $\phi_f(r,0)=\phi_{f,0}$, so that the initial state is the reference state and this is consistent with the boundary conditions.

\subsubsection{Driving conditions}\label{driving}

For $t>0$, we assume that fluid is injected from the origin at either a fixed driving pressure difference $\Delta{p}$ or a fixed flow rate $q$. These quantities are related by
\begin{equation}\label{q_dp_relation}
    q=\frac{\Delta{p}}{\displaystyle \int_a^b\displaystyle \frac{1}{kr}\,\mathrm{d}r} +\frac{\displaystyle \int_a^b\displaystyle \frac{v_s}{k}\,\mathrm{d}r}{\displaystyle \int_a^b\displaystyle \frac{1}{kr}\,\mathrm{d}r},
\end{equation}
which is derived from Equation~(\ref{Vel}). Note that the steady state response of the cylinder to an imposed value of $q$ corresponds to some \textit{a priori} unknown value of $\Delta{p}$. As a result, the same steady state can be achieved by instead imposing this value of $\Delta{p}$. However, the transient evolution to this identical steady state will be very different (\textit{c.f.,} \S\ref{secdpvsq}).

\section{Model set up and solution methods}\label{fp}

\subsection{Governing equations}

The conservation laws for the three classes of models, Equation~(\ref{govern}) for the Q and N models and Equation (\ref{eq:phi_to_ulin}b) for the L models, are written terms of $\phi_f$ and $p$. For the L models, it is straightforward to rewrite this as a partial differential equation (PDE) in $\phi_f$ using Equations~(\ref{eq:phi_to_ulin}a) and (\ref{dpdrlin}). For the Q and N models, however, this process is much less straightforward. Additionally, the boundary conditions for all models are written in terms of $u_s$ and $\sigma^\prime_r$, the latter being readily expressible in terms of $u_s$. To rewrite all of these problems as closed initial boundary value problems (IBVPs) in terms of a single dependent variable, we therefore use $u_s$ in lieu of $\phi_f$.

\subsubsection{L models}

Equations~(\ref{dpdrlin}) and (\ref{eq:phi_to_ulin}) yield
\begin{subequations}
\begin{equation}\label{Lmodel}
    \frac{\partial{u_s}}{\partial{t}}= v_s=\frac{q}{r}+k[u_s]\frac{\partial}{\partial{r}} \left[\frac{1}{r}\frac{\partial}{\partial{r}}(ru_s)\right],
\end{equation}
where
\begin{equation}\label{k}
    k[u_s]\equiv1 \quad 
    \text{or} \quad
    k[u_s] = \frac{\left[(1-\phi_{f,0})\left(\frac{\partial{u_s}}{\partial{r}} +\frac{u_s}{r}\right)+\phi_{f,0}\right]^3}{\phi_{f,0}^3\left(1-\frac{\partial{u_s}}{\partial{r}}-\frac{u_s}{r}\right)^2}
\end{equation}
\end{subequations}
for the L-$k_0$ or L-$k_\mathrm{KC}$ models, respectively. Subjected to $B^a[u_s]=B_i^b[u_s]=0$ for $i=1,2$, this now constitutes a closed IBVP. For the L-$k_0$ model driven by a fixed flux $q$, this problem is analytically tractable and we present solutions below (\textit{c.f.,} \S\ref{ana_zd} and Appendix~A).

\subsubsection{Q and N models} 

The conservation law for the Q and N models is
\begin{subequations}\label{QNgov}
\begin{equation}\label{gen}
    \frac{\partial{u_s}}{\partial{t}}=\frac{1}{\lambda_r}v_s= \left(1-\frac{\partial{u_s}}{\partial{r}}\right)\left[\frac{q}{r}+k[u_s]\frac{\partial{p}}{\partial{r}}\right],
\end{equation}
where
\begin{equation}
    k[u_s]\equiv{}1 \quad
    \text{or} \quad
    k[u_s]=\frac{\left[(1-\phi_{f,0})\left(\frac{\partial{u_s}}{\partial{r}} +\frac{u_s}{r}-\frac{u_s}{r}\frac{\partial{u_s}}{\partial{r}}\right)+\phi_{f,0}\right]^3}{\phi_{f,0}^3\left(1-\frac{\partial{u_s}}{\partial{r}}-\frac{u_s}{r}+\frac{u_s}{r}\frac{\partial{u_s}}{\partial{r}}\right)^2}
\end{equation}
\end{subequations}
for the `-$k_0$' or `-$k_{\mathrm{KC}}$' models, respectively. Note that it can be shown that Equation~(\ref{gen}) is equivalent to Equation~(\ref{govern}). For the Q models, we pair Equation~(\ref{QNgov}) with Equation~(\ref{dpdrlin}) to obtain a PDE in terms of $u_s$. For the N models, we pair Equation~(\ref{QNgov}) with Equation~(\ref{dpdrhenck}). Both of these cases, subject to $B^a[u_s]=B^b_i[u_s]=0$ for $i=1,2$, form the corresponding closed IBVPs.

\subsection{Analytical solutions: L-$k_0$ for fixed $q$}\label{ana_zd}

We now develop analytical solutions for the L-$k_0$ model for a fixed driving flow rate $q$. For brevity, we include the solution for the constrained cylinder below and that for the cylinder subject to a fixed radial effective stress at the outer boundary in Appendix~A.

The IBVP for the constrained cylinder is
\begin{subequations}\label{IBVP}
\begin{align}
    \frac{\partial{u_s}}{\partial{t}}-\frac{\partial}{\partial{r}} \left[\frac{1}{r}\frac{\partial}{\partial{r}}(ru_s)\right] =\frac{q}{r} & \qquad a_0<r<1, \quad t>0  \label{TL} \\
    u_s(r,0)=0 &\qquad a_0<r<1, \quad t=0 \label{IC} \\
    B^{a_0}[u_s]=\frac{\partial{u_s}}{\partial{r}} +\Gamma\frac{u_s}{a_0}=0 & \qquad r=a_0, \qquad \ \ \ t>0 \label{BC1} \\
    B^{b_0}_1[u_s]=u_s(1,t)=0 & \qquad r=1, \qquad \ \quad t>0 \label{BC2}.
\end{align}
\end{subequations} 
We approach this problem using separation of variables for a non-homogeneous PDE. We begin by substituting the separable ansatz
\begin{equation}\label{ansatz1}
    u_s = \mathcal{T}(t)\mathcal{R}(r)
\end{equation}
into the associated homogeneous PDE
\begin{equation}\label{homo1}
    \frac{\partial{u_s}}{\partial{t}}-\frac{1}{r} \left[\frac{\partial}{\partial{r}}\left(r\frac{\partial{u_s}}{\partial{r}}\right)-\frac{u_s}{r}\right]=0
\end{equation}
to obtain
\begin{equation}
    \frac{\mathcal{T}^\prime(t)}{\mathcal{T}(t)} =\frac{\left[r\mathcal{R}^\prime(r)\right]^\prime-\mathcal{R}(r)/r}{r\mathcal{R}(r)}= -\omega^2
\end{equation}
for some constant $\omega$. The spatial problem is thus
\begin{subequations}\label{eigenVP}
\begin{equation}\label{SL}
    \left[r\mathcal{R}^\prime(r)\right]^\prime - \frac{\mathcal{R}(r)}{r} +r\omega^2 \mathcal{R}(r)= 0
\end{equation}
\begin{equation}\label{BCS}
    \mathcal{R}^\prime(a_0)+\Gamma\frac{\mathcal{R}(a_0)}{a_0} = \mathcal{R}(1) = 0,
\end{equation}
\end{subequations} 
which constitutes a Sturm-Liouville eigenvalue problem. This has two key advantageous properties: Firstly, there exist infinitely many strictly increasing real eigenvalues; secondly, the corresponding eigenfunctions constitute an orthogonal basis \citep{pinchover2005introduction}. The solution to this eigenvalue problem can, therefore, be used to construct solutions to the original IBVP (\ref{IBVP}) in terms of an infinite series of eigenfunctions.

The boundary conditions are homogeneous (Robin at $r=a_0$ and Dirichlet at $r=1$), allowing us to solve the spatial problem (\ref{eigenVP}) in its current form. We rewrite Equation~(\ref{SL}) as
\begin{equation}
    \mathcal{R}''(r)+\frac{\mathcal{R}'(r)}{r}+\mathcal{R}(r)\left(\omega^2-\frac{1}{r^2}\right) = 0,
\end{equation}
recognising this as Bessel's differential equation with solution
\begin{equation}
    \mathcal{R}(r)=c_1J_1(\omega{}r)+c_2Y_1(\omega{}r),
\end{equation}
where $c_1$ and $c_2$ are constants to be determined and $J_\nu$ and $Y_\nu$ are Bessel functions of the first and second kind, respectively, of order~$\nu$. Using Equation (\ref{BCS}), we then obtain
\begin{multline}\label{eq:2x2aa}
    \left(\begin{array}{cc}
        J_1(\omega) & Y_1(\omega) \\
        \frac{\omega}{2}[J_0(\omega{}a_0)-J_2(\omega{}a_0)] +\frac{\Gamma}{a_0}J_1(\omega{}a_0) &
        \frac{\omega}{2}[Y_0(\omega{}a_0)-Y_2(\omega{}a_0)] +\frac{\Gamma}{a_0}Y_1(\omega{}a_0) \\
    \end{array} \right) \\
    \left(\begin{array}{c}
        c_1 \\
        c_2 \\
    \end{array} \right) =
    \left(\begin{array}{c}
        0 \\
        0 \\
    \end{array} \right).
\end{multline}
For non-trivial solutions, we set the determinant of the matrix to zero, 
\begin{multline}\label{det}
    J_1(\omega)\left\{\frac{\omega}{2} \left[Y_0(\omega{}a_0)-Y_2(\omega{}a_0)\right]+\frac{\Gamma}{a_0}Y_1(\omega{}a_0)\right\}- \\
    Y_1(\omega)\left\{\frac{\omega}{2} \left[J_0(\omega{}a_0)-J_2(\omega{}a_0)\right]+\frac{\Gamma}{a_0}J_1(\omega{}a_0)\right\} = 0,
\end{multline}
which, by the Freedholm Alternate Theorem, provides an infinite number of solutions for the infinite set of distinct eigenvalues $\omega = \omega_n$ that satisfy Equation (\ref{det}). For the boundary condition at $\mathcal{R}(1)$, we have that
\begin{equation}
    c_2 = -c_1\frac{J_1(\omega_n)}{Y_1(\omega_n)},
\end{equation}
and we further take $c_1=1$ without loss of generality. We then have an infinite number of eigenfunctions $\mathcal{R}_n(r)$ given by
\begin{equation}
    \mathcal{R}_n(r) = J_1(\omega_n{}r)-\frac{J_1(\omega_n)}{Y_1(\omega_n)} Y_1(\omega_n{}r).
\end{equation}
Note that, as the $\omega_n$ are eigenvalues of a Sturm-Liouville problem, and by construction are positive, they must also satisfy
\begin{equation}
    0<\omega_1<\omega_2<\cdots \quad \mathrm{with} \quad \omega_n \to \infty \quad \mathrm{as} \quad n\to \infty, 
\end{equation}
The eigenfunctions are orthogonal with respect to a weighted inner product with associated weighting function $r$ (the coefficient of $\omega^2 \mathcal{R}$ in Equation (\ref{SL})). The weighted inner product is therefore given by
\begin{subequations}\label{properties}
\begin{equation}
    \left\langle f,g \right\rangle=\int_{a_0}^1 r f(r)g(r)\,\mathrm{d}r,
\end{equation}
so that
\begin{equation}
    \frac{\langle \mathcal{R}_n,\mathcal{R}_m\rangle}{\langle \mathcal{R}_n,\mathcal{R}_n\rangle} = 
    \left\{\begin{array}{ll}
        0 & \mbox{if $n\neq m$};\\
        1 & \mbox{if $n=m$}.
    \end{array} \right.
\end{equation}
\end{subequations}
We must now determine the corresponding $\mathcal{T}_n(t)$ such that 
\begin{equation}\label{asdfd}
    u_s = \sum_{n=1}^{\infty}\mathcal{T}_n(t)\mathcal{R}_n(r)
\end{equation}
satisfies the original IBVP (\ref{IBVP}), despite the fact that our initial separable ansatz was applied to the homogeneous problem. Substituting Equation~(\ref{asdfd}) into Equation~(\ref{TL}), we obtain
\begin{equation}\label{q_decompose_step1}
    \frac{q}{r} = \frac{\partial{u_s}}{\partial{t}} - \frac{\partial }{\partial{r}}\left[\frac{1}{r}\frac{\partial}{\partial{r}}(u_sr)\right] =\sum_{n=1}^{\infty}\left[\mathcal{T}'_n(t)+\omega_n^2\mathcal{T}_n(t)\right] \mathcal{R}_n(r).
\end{equation}
This motivates decomposing $q/r$ in terms of an infinite series of eigenfunctions,
\begin{equation}\label{q_decompose_step2}
    \frac{q}{r} = \sum_{n=1}^{\infty} f_n \mathcal{R}_n(r),
\end{equation}
for some infinite set of constants $f_n$. Combining Equations~(\ref{q_decompose_step1}) and (\ref{q_decompose_step2}) leads to the ordinary differential equation (ODE)
\begin{equation}
    \mathcal{T}'_n(t)+\omega_n^2\mathcal{T}_n(t)=f_n,
\end{equation}
which has solution
\begin{equation}
    \mathcal{T}_n(t)=\frac{f_n}{\omega_n^2}\left(1-e^{-\omega_n^2t}\right).
\end{equation}
We evaluate $f_n$ from Equation~(\ref{q_decompose_step2}) by invoking (\ref{properties}), yielding
\begin{equation}\label{simples}
    f_n = \frac{q\int_{a_0}^1\mathcal{R}_n(r)\,\mathrm{d}r} {\int_{a_0}^1r[\mathcal{R}_n(r)]^2\,\mathrm{d}r} =\frac{q\left[Y_1(\omega_n)\right]^2I_0}{\left[Y_1(\omega_n)\right]^2I_1 +\left[J_1(\omega_n)\right]^2I_2 - 2J_1(\omega_n)Y_1(\omega_n)I_3 },
\end{equation}
where
\begin{subequations}\label{Ints}
\begin{equation}
    I_0 \defeq \frac{\Big[J_1(\omega_n)Y_0(\omega_n{}r)-Y_1(\omega_n)J_0(\omega_n{}r)\Big]^1_{a_0}}{\omega_nY_1(\omega_n)}, 
\end{equation}
\begin{equation}
    I_1 \defeq \left[\frac{r^2}{2}\Big\{J_1(\omega_n{}r)^2-J_0(\omega_n{}r)J_2(\omega_n{}r)\Big\}\right]^1_{a_0}, 
   \end{equation}
\begin{equation}
    I_2 \defeq \left[\frac{r^2}{2}\Big\{Y_1(\omega_n{}r)^2-Y_0(\omega_n{}r)Y_2(\omega_n{}r)\Big\}\right]^1_{a_0}, 
\end{equation}
and 
\begin{equation}
    I_3 \defeq \left[\frac{r^2}{2}\Big\{J_0(\omega_n{}r)Y_0(\omega_n{}r)+J_1(\omega_n{}r)Y_1(\omega_n{}r)\Big\} -\frac{r}{\omega_n}\Big\{J_0(\omega_n{}r)Y_1(\omega_n{}r)\Big\}\right]^1_{a_0}.
\end{equation}
\end{subequations}
Finally, the solution is given by
\begin{equation}\label{anasolconstrained}
    u_s = \sum_{n=1}^{\infty} \left[\frac{f_n}{\omega_n^2}\left(1-e^{-\omega_n^2{}t}\right)\right] \left[J_1(\omega_n{}r)-\frac{J_1(\omega_n)}{Y_1(\omega_n)} Y_1(\omega_n{}r)\right],
\end{equation}
where the $\omega_n$ are the solutions to Equation (\ref{det}) and the $f_n$ are defined by Equations (\ref{simples}--\ref{Ints}). 

\subsection{Numerical solution method}

We solve all of the nonlinear problems, as well as the L-$k_0$ model for fixed $\Delta{p}$, numerically. We do so by extending the method presented in \citet{auton2017arteries} by combining Chebyshev spectral collocation with the method of lines. That is, we discretise the spatial domain into the $N$ Chebyshev points and then approximate spatial derivatives using a Chebyshev differentiation matrix \cite{auton2017arteries}. We then integrate the resulting system of differential algebraic equations (DAEs) in time with \verb+MATLAB+ using \verb+ode15s+ \cite{piche2007solving, piche2009matrix}. This pseudospectral method has proven to be more robust than classical finite volumes or finite differences, as well as more convenient for handling certain combinations of boundary and driving conditions (\textit{e.g.}, fixed $\Delta{p}$ with fixed radial effective stress at the outer boundary).

\subsubsection{Fixed $q$}\label{q}

We next outline the implementation of the numerical scheme for fixed $q$. We denote the general form of Equations (\ref{Lmodel}) and (\ref{gen}) via
\begin{equation}\label{GenFmq}
    M[u_s]\frac{\partial{u_s}}{\partial{t}} = {F}[u_s] + {c}[u_s]\frac{q}{r},
\end{equation}
where $M$, $F$, and $c$ are continuous partial-differential operators in $r$. In Appendix~B, we provide expression for $M$, $F$, and $c$ for all models and boundary conditions for fixed $q$. Following the method of lines, we discretise $u_s$ and $r$ in space and, using first and second order Chebyshev differentiation matrices of size $N \times N$, we discretise the operators $M$, $F$, and $c$. Below, we denote spatially discretised quantities with hats; vectors and matrices are additionally in bold.

For $a<r<b$, we then have a system of coupled ODEs in time. At $r=a$ and $r=b$, we enforce the spatially discretised boundary conditions, $\hat{B}^{a}(\boldsymbol{\hat{u}}) =\hat{B}^b_i(\boldsymbol{\hat{u}})= 0$ for $i =1,2$ (\textit{c.f.,} Equation (\ref{bclucy})), which are algebraic in $t$. Together, these equations constitute a system of DAEs. At each time step, the domain deforms and we use Equations (\ref{innerKBC}) and (\ref{outerKBC}) as appropriate to update the grid. It is clear that $\boldsymbol{\hat{F}}(\boldsymbol{\hat{u}})$, $\boldsymbol{\hat{c}}(\boldsymbol{\hat{u}})$, and thus the right-hand side of Equation (\ref{GenFmq}), are all vectors of length $N$. To enforce the boundary conditions, the first and last entries of the right-hand side must be $\hat{B}^{a}(\boldsymbol{\hat{u}})$ and $\hat{B}^b_i(\boldsymbol{\hat{u}})$, respectively. We express this system of DAEs using a mass matrix $\boldsymbol{\hat{M}}(\boldsymbol{\hat{u}})$, which is the Chebyshev spatial discretisation of $M[u_s]$. The mass matrix pre-multiplies the time derivative $\partial{u_s}/\partial{t}$ and enables us to enforce the boundary conditions ($\hat{B}^{a}(\boldsymbol{\hat{u}}) =\hat{B}^b_i(\boldsymbol{\hat{u}})= 0$) by setting the first and last rows of $\boldsymbol{\hat{M}}(\boldsymbol{\hat{u}})$ identically equal to zero.

We integrate this system of DAEs in time in \verb+MATLAB+ using \verb+ode15s+. When tractable, we provide the solver with an analytical Jacobian,
\begin{equation}
    \frac{\mathrm{d}}{\mathrm{d} \boldsymbol{\hat{u}}} \left(\boldsymbol{\hat{F}}(\boldsymbol{\hat{u}})+\boldsymbol{\hat{c}}(\boldsymbol{\hat{u}})\frac{q}{ \boldsymbol{\hat{r}}}\right),
\end{equation}
accounting appropriately for the moving boundaries. Note that, as the problem becomes more nonlinear, it is numerically beneficial to initialise the problem with a flow rate of zero and ramp this to the desired value of $q$ over a short time (at most $10^{-5}$). It is clear from the results that this does not impact the transient evolution for times greater than the ramping time.

\subsubsection{Fixed $\Delta{p}$}\label{dp}

Driving with a fixed pressure difference $\Delta{p}$ leads to an integro-PDE as $q$ appears explicitly in the PDE (\ref{GenFmq}) and $q$ is related to $\Delta{p}$ via Equation (\ref{q_dp_relation}), which contains a spatial integral of $\partial{u_s}/\partial{t}$. In our Chebyshev framework, we discretise these integrals using Lobatto's integral formula, which allows us to incorporate them into the system of DAEs. Lobatto's integral formula states that
\begin{equation}\label{lobatto}
    \int_a^b g(r) \, \mathrm{d}r = \frac{\pi}{N-1}\left(\frac{b-a}{2}\right) \sum_{k=1}^N (1-\hat{X}_k^2)^\frac{1} {2}\hat{g}_k + Res_N
\end{equation}
for some function $g(r)$, $r \in [a,b]$, where $\hat{X}_k\in[-1,1]$ is the $k^\mathrm{th}$ Chebyshev point. We neglect the residual $Res_N$, which we assume decays rapidly as $N\to{}\infty$ \cite{bjornaraa2013pseudospectral, abramowitz1964handbook}.

Rewriting Equation (\ref{GenFmq}) using Equation (\ref{q_dp_relation}) gives 
\begin{equation}\label{GeneralForm}
    M[u_s]\frac{\partial{u_s}}{\partial{t}} = {F}[u_s] + c[u_s]\left(\frac{\Delta{p}}{r\int_a^b\frac{1}{kr}\,\mathrm{d}r} +\frac{\int_a^b\frac{v_s}{k}\,\mathrm{d}r}{r\int_a^b\frac{1}{kr}\,\mathrm{d}r},\right).
\end{equation}
The solid velocity $v_s$ is related to $\partial{u_s}/\partial{t}$ via Equation (\ref{eq:phi_to_ulin}) for the L models and Equation (\ref{Vel}) for the Q and N models, and we again discretise these relations in terms of $\boldsymbol{\hat{u}}$, $\boldsymbol{\hat{r}}$, and the Chebyshev differentiation matrices. Using Equation (\ref{lobatto}), we then express the relevant integral in Equation (\ref{GeneralForm}) as a finite sum of $\mathrm{d}{\hat{\boldsymbol{u}}}/\mathrm{d}{t}$. Similarly, the other integral in Equation (\ref{GeneralForm}) is readily approximated with Lobatto's integral formula. We incorporate the resulting sum of $\mathrm{d}{\hat{\boldsymbol{u}}}/\mathrm{d}{t}$ into a new mass matrix $\boldsymbol{\hat{m}}$, whose first and last rows are once again identically equal to zero. We then write the fully discretised system in the form $\boldsymbol{\hat{m}}(\boldsymbol{\hat{u}}) {\mathrm{d}{\boldsymbol{\hat{u}}}}/{\mathrm{d}{t}} = \boldsymbol{\hat{f}} (\boldsymbol{\hat{u}})$, where $\boldsymbol{\hat{f}}(\boldsymbol{\hat{u}})$ is a vector of length $N$ whose first and last entries are $\hat{B}^{a}(\boldsymbol{\hat{u}})$ and $\hat{B}^b_i(\boldsymbol{\hat{u}})$, respectively.

We use this method for all of the fixed $\Delta{p}$ cases. As the problem becomes more nonlinear, $\boldsymbol{\hat{m}}$ becomes more complicated and it is again numerically beneficial to ramp $\Delta{p}$ from zero to the desired value over a short time (at most $10^{-5}$).

\section{Results}\label{results}

We now consider the transient evolution of the deformation for different driving method (fixed $\Delta{p}$ \textit{vs.} fixed $q$), confinement (confined \textit{vs.} unconfined), and geometry (varying $a_0$) for six model combinations: L-$k_0$, L-$k_\mathrm{KC}$, Q-$k_0$, Q-$k_\mathrm{KC}$, N-$k_0$, and N-$k_\mathrm{KC}$.

\subsection{Fixed $\Delta{p}$ \textit{vs.} fixed $q$ for constrained thick-walled cylinders}\label{secdpvsq}

\begin{figure}[t]
    \centering
    \includegraphics[width=5in]{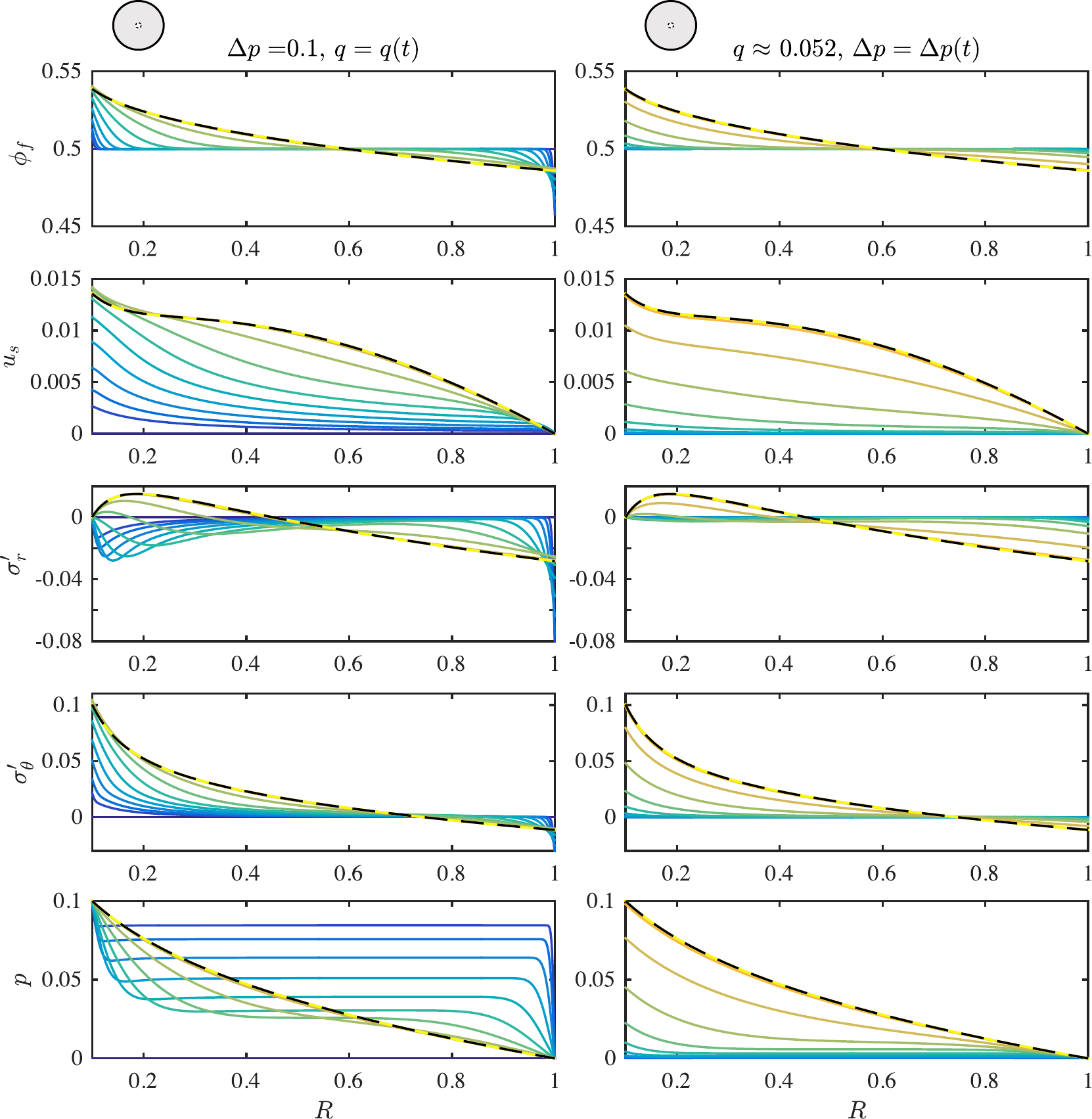}
    \caption{The time evolution of a constrained thick-walled cylinder ($a_0=0.1$) for the Q-$k_\mathrm{KC}$ model, where the flow is driven by fixed $\Delta{p}=0.1$ (left column) or by the fixed $q\approx 0.052$ that leads to the same steady state (right column). We show the solution at twelve times, logarithmically spaced from $t = 10^{-5}$ (blue) to $t = 2.5$ (yellow). We also include the initial condition and the steady state for reference (solid and dashed black lines, respectively). We plot all quantities against the Lagrangian radial coordinate $R\defeq{}r(t)-u_s(r,t)$. For this scenario, the evolution to the common steady state occurs much more quickly for fixed $\Delta{p}$ than for fixed $q$. Additionally, some degree of non-monotonicity in time is evident in every quantity for fixed $\Delta{p}$, whereas all quantities except for $\sigma_r^\prime$ evolve monotonically for fixed $q$. \label{all_quants_dp_vs_q} }
\end{figure}

\begin{figure}[tb]
    \centering
    \includegraphics[width=5in]{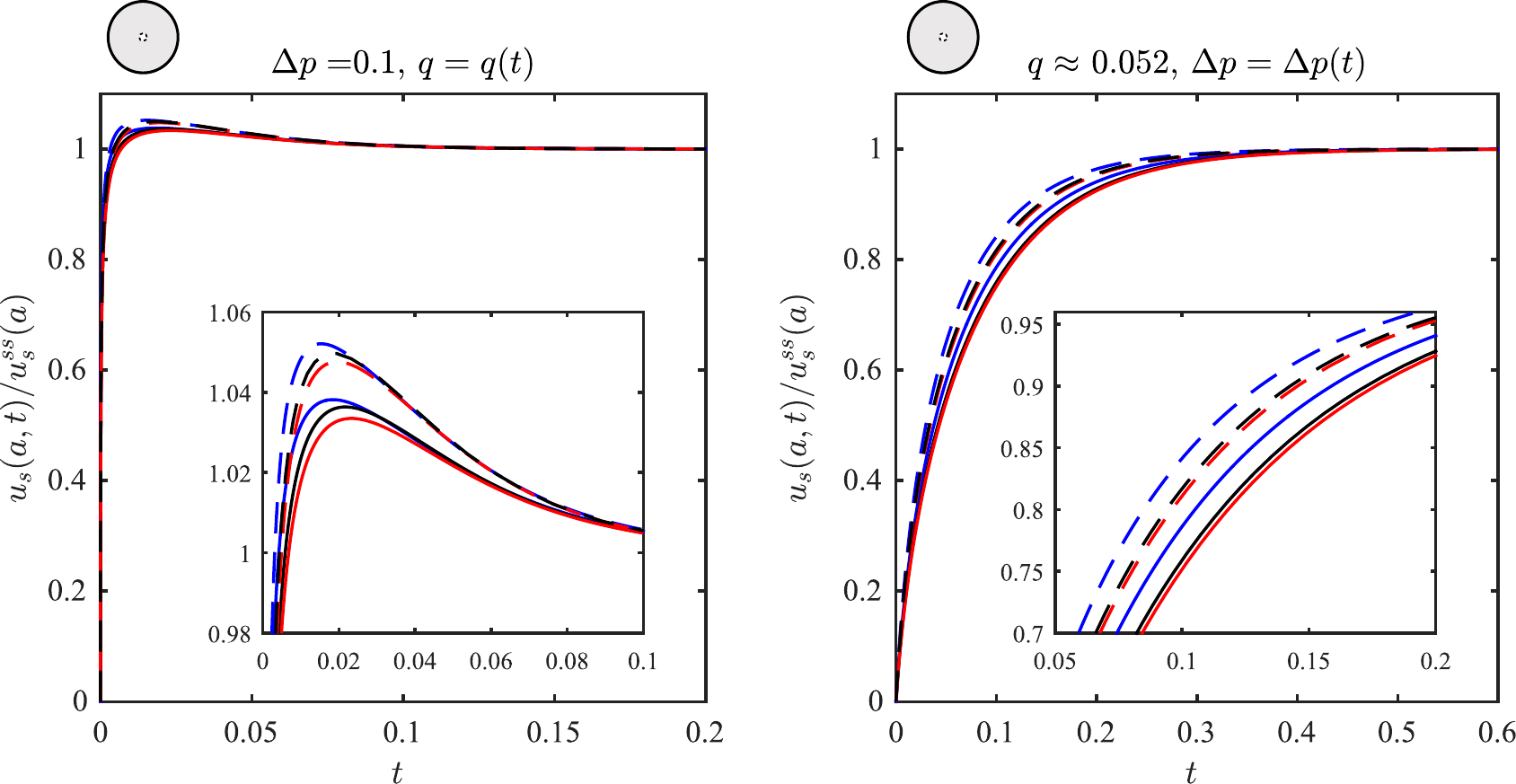}
    \caption{The time evolution of the normalised displacement at the inner boundary for all six models for the same scenario and parameter values as in Figure \ref{all_quants_dp_vs_q}. We plot the L models (blue), Q models (black), and N models (red) for both constant permeability (solid lines) and Kozeny-Carman permeability (dashed lines). The insets highlight the differences between models at early times. All models exhibit the same qualitative behaviour, and the deformation evolves much more quickly in the fixed $\Delta{p}$ case than in the fixed $q$ case. In the fixed $\Delta{p}$ case, nonlinear kinematics and nonlinear elasticity moderate the relative overshoot, whereas Kozeny-Carman permeability exacerbates it. We present the same figure but for the normalised azimuthal effective stress in Appendix~C (Figure~C1). \label{dpvsq_u_a} }
\end{figure}

\begin{figure}[tb]
    \centering
    \includegraphics[width=5in]{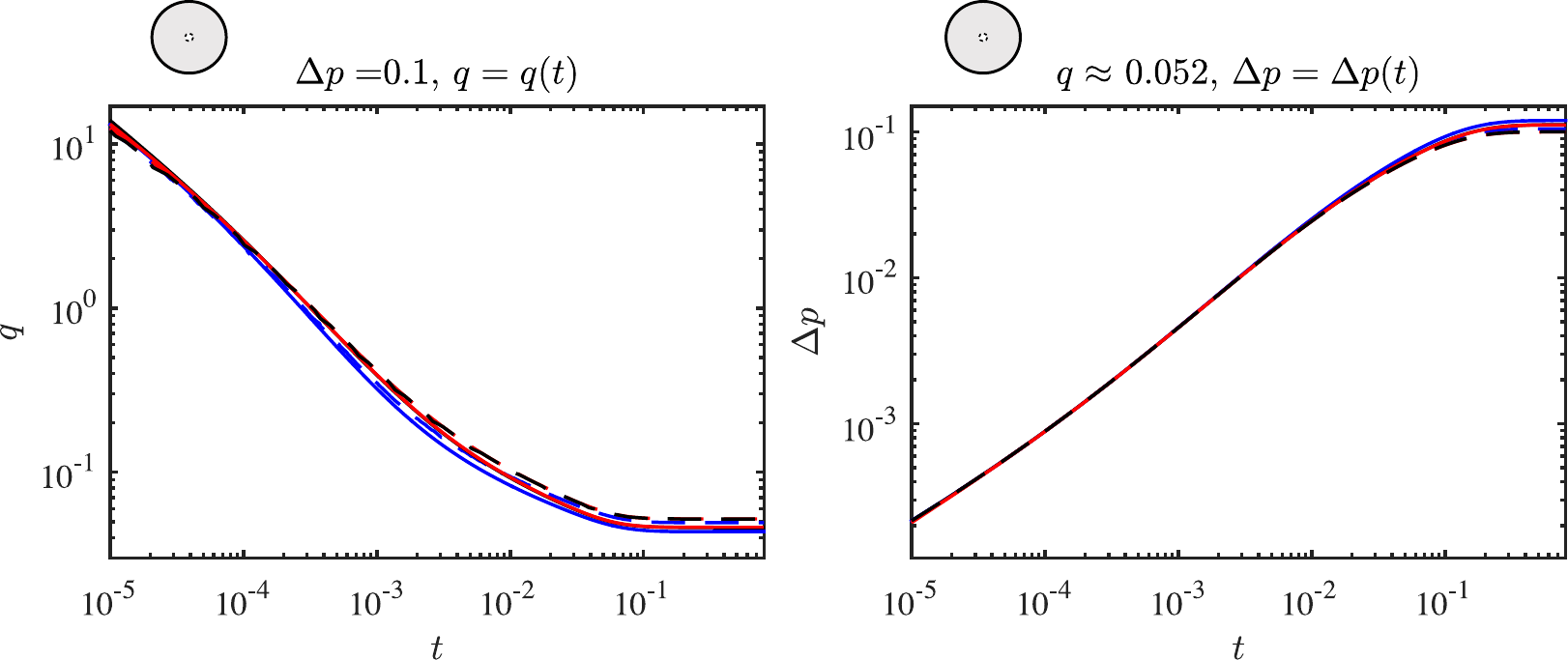}
    \caption{The time evolution of $q$ for fixed $\Delta{p}$ (left), and of $\Delta{p}$ for fixed $q$ (right), for all 6 models and for the same scenario and parameter values as in Figures \ref{all_quants_dp_vs_q} and \ref{dpvsq_u_a}. Line colours and styles are also the same as in Figure \ref{dpvsq_u_a}. For fixed $\Delta{p}$, $q(t)$ is initially very large and decreases towards steady state; for fixed $q$, $\Delta{p}(t)$ is initially very small and gradually increases towards steady state. These opposite evolutions result from the same physics. \label{qvsdp} }
\end{figure}

\begin{figure}[tb]
    \centering
    \includegraphics[width=5in]{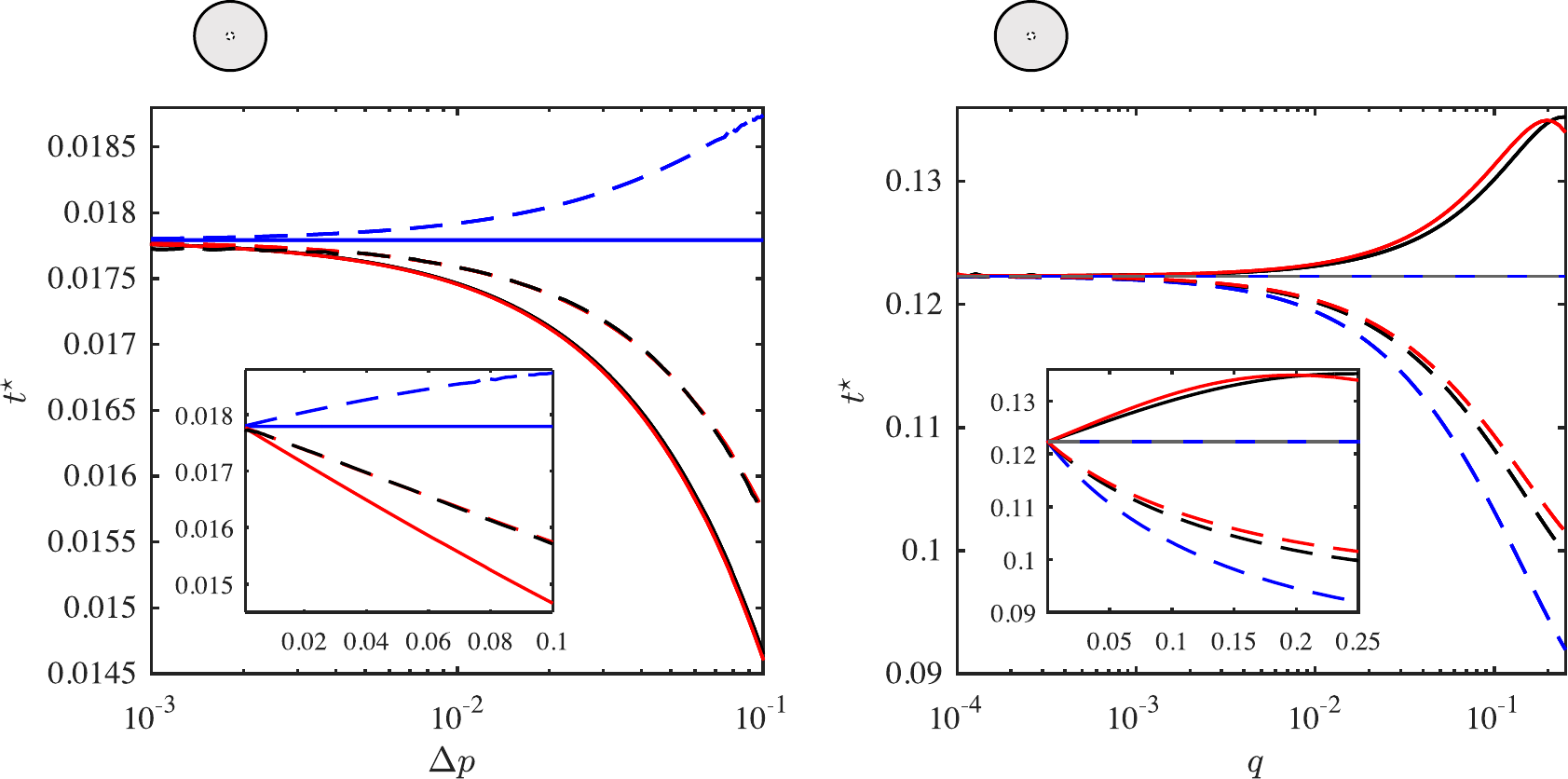}
    \caption{The evolution timescale $t^\star$ normalised by the evolution timescale for the L-$k_0$ model of the constrained thick-walled cylinder ($a_0=0.1$) for all six models as a function of $\Delta{p}$ (left), and of $q$ (right). Line colours and styles are the same as in Figures~\ref{dpvsq_u_a}--\ref{qvsdp}, we additionally show the analytical dynamic L-$k_0$ fixed $q$ solution (right, dashed grey lines). For the L-$k_0$ model, $t^\star$ is independent of driving strength in both cases. For the Q and N models, $t^\star$ is very similar. For the fixed $\Delta{p}$ cases, $t^\star$ is roughly one order of magnitude smaller than for the fixed $q$ cases. To highlight the extent to which nonlinearity accelerates or decelerates the transient evolution in each case relative to the L-$k_0$ model, we plot these same results as $t^\star/t^\star_{\mathrm{L}k0}$ against $q$ and $\Delta{p}$ in Appendix~E (Figure~E1). \label{qvsdp_tch} }
\end{figure}

We begin by considering the impact of driving method for the constrained thick-walled cylinder ($a_0=0.1$). To do so, we solve the problem for a fixed $\Delta{p}$ and then again for the fixed value of $q$ that results in the same steady. We consider the time evolution of the deformation for this scenario in Figures~\ref{all_quants_dp_vs_q}--\ref{qvsdp}. In Figure~\ref{all_quants_dp_vs_q}, we show the time evolution of all key quantities for the Q-$k_\mathrm{KC}$ model. In Figure~\ref{dpvsq_u_a}, we examine the impact of model choice by considering the evolution of the displacement at the inner boundary for all six models. In Figure~\ref{qvsdp}, we examine the time evolution of $q$ when $\Delta{p}$ is fixed, and of $\Delta{p}$ when $q$ is fixed, again for all six models. In Figure~\ref{qvsdp_tch}, we consider the characteristic evolution timescale for this geometry and boundary condition over a wide range of $\Delta{p}$ and $q$ values, again for all six models.

In Figure~\ref{all_quants_dp_vs_q}, we plot the evolution of all key quantities to steady state for the Q-$k_\mathrm{KC}$ model for fixed $\Delta{p}=0.1$ (left column), and for the fixed $q \approx 0.052$ that leads to the same steady state (right column).
For fixed $\Delta{p}$, $\phi_f$ (first row), $u_s$ (second row), and $\sigma^\prime_\theta$ (fourth row) show similar qualitative behaviours: for the majority of the Lagrangian radius $R\defeq{}r(t)-u_s(r,t)$, these quantities evolve monotonically towards their steady-state values. Additionally, all of these quantities overshoot their steady-state values near the inner and outer boundaries. Near $R=a_0$, they overshoot once before relaxing towards their steady states; near $R=1$, they overshoot once at early times and then again at intermediate times before relaxing towards their steady states. The cylinder effectively ``over-deforms'' near both the free inner boundary and the confined outer boundary; as time progresses, this deformation relaxes. This overshoot implies that the largest stresses (\textit{e.g.}, the maximum value of $\sigma_\theta^\prime$) occur at some intermediate time, rather than at steady state, which has implications for problems concerning material failure such as hydraulic fracturing. Note, however, that this overshoot does not occur for all values of $a_0$. The radial effective stress $\sigma_r^\prime$ (third row) mirrors this behaviour near $R=1$. Near $R=a_0$, however, $\sigma_r^\prime$ initially decreases into strong compression before increasing to its tensile steady-state value. The pressure $p$ has a fixed value at both boundaries by construction, $p(a)\equiv{}\Delta{p}$ and $p(b)\equiv{}0$, and exhibits an approximately uniform interior value bracketed by sharp boundary layers at $R=a_0$ and $R=1$ that spread with time, which is consistent with classical consolidation theory. These boundary layers imply that the deformation is initially focused near the boundaries, which is linked to the overshoot in deformation through mechanical equilibrium (\textit{c.f.}, Eq.~\ref{dpdrlin} and \S4c of \citet{auton2017arteries}).

For fixed $q$, in contrast, the evolution is much slower and all quantities except for $\sigma_r^\prime$ evolve monotonically in time. The radial effective stress $\sigma_r^\prime$ decreases initially, such that it becomes strictly non-positive for some time before evolving towards its steady state by increasing into tension near the inner boundary and by decreasing further into compression near the outer boundary. The slower evolution timescale is a result of the fact that driving with fixed $\Delta{p}$ leads to an initially large value of $q$ that decreases towards its steady-state value, driving the material very aggressively at early times (\textit{c.f.,} Figure~\ref{qvsdp}, left); driving with fixed $q$, in contrast, leads to a initially small value of $\Delta{p}$ that increases to steady state, driving the material more gently at early times (\textit{c.f.,} Figure~\ref{qvsdp}, right).

In Figure~\ref{dpvsq_u_a}, we consider the normalised displacement at the inner boundary $\bar{u}_a \defeq u_s(a,t)/u_{s}^\mathrm{ss}(a)$ for all six models, where $u_{s}^\mathrm{ss}$ denotes the displacement at steady state such that $\bar{u}_a \to{}1$ as $t \to\infty$. Note that we compare each model to \textit{its own} steady state, and the six steady states are not the same~\citep{auton2017arteries}. We also compare the evolution of $\bar{u}_a$ for fixed $\Delta{p}$ (left) \textit{vs.} fixed $q$ (right). For fixed $\Delta{p}$, all six models exhibit the overshoot noted above. This overshoot is most pronounced in the -$k_\mathrm{KC}$ models. Within each permeability grouping, the L models show the most relative overshoot and the N models the least, implying that both rigorous kinematics and nonlinear elasticity moderate the relative overshoot, while deformation-dependent permeability exacerbates it. For fixed $q$, the ordering of the models is the same as for fixed $\Delta{p}$, but the evolution timescale is clearly much slower. These results suggest that, for a constrained thick walled cylinder, driving method has much stronger impact than model choice on both the qualitative nature and the timescale of the transient evolution.

In Figure~\ref{qvsdp}, we examine the time evolution of $q$ for fixed $\Delta{p}$ (left), and of $\Delta{p}$ for fixed $q$ (right). For fixed $\Delta{p}$, $q(t)$ is initially very large and then decreases towards steady state; this behaviour is qualitatively the same for all models. The flow rate $q(t)$ decreases with time because the solid moves radially outward at early times, so a relatively large fluid velocity is needed to generate the required pressure difference. The solid slows over time as deformation increasingly resists further motion, and the fluid velocity slows accordingly. This leads to an initially large flow rate that decays towards the steady-state value, for which the solid is stationary. For fixed $q$, the reverse occurs: The solid and the fluid both contribute to the fixed total flux at early times, moving together such that only a relatively small value of $\Delta{p}$ is needed. As the solid slows, the relative velocity of the fluid must increase to preserve the fixed total flux and the necessary $\Delta{p}$ increases. At steady state, the solid is stationary, the fluid provides the entire flux, and $\Delta{p}$ is largest.

In Figure~\ref{qvsdp_tch}, we consider the characteristic evolution timescale of the various models shown in Figures~\ref{all_quants_dp_vs_q}--\ref{qvsdp}. We define the evolution timescale as the time $t^\star$ at which the Euclidean norm of the relative difference from steady state is equal to 0.2, where we measure the relative difference from steady state as $[u_s^{ss}(r)-u_s(r,t^\star)]/u_s^{ss}(r)$ and we generate the steady state as in \citet{auton2017arteries}. Note that this metric is weakly influenced quantitatively by the fact that our solutions are defined on a Chebyshev grid, but this is consistent across all cases and makes no qualitative difference in the results.

For the L-$k_0$ model, the timescale is independent of driving strength for both fixed $\Delta{p}$ and fixed $q$. This is due to our definition of $t^\star$ in terms the relative difference from steady state and the fact that, for the L-$k_0$ model, $u_s(r,t)$ and $u_s^{ss}(r)$ are proportional to $\Delta{p}$ for fixed $\Delta{p}$, and proportional to $q$ for fixed $q$. In the latter case, this is obvious from the analytical solutions for the constrained and unconstrained cylinders. Although $t^\star$ is independent of driving strength for the L-$k_0$ model, the driving method, boundary condition, and geometry (\textit{i.e.}, $a_0$) all have significant impacts on the timescale (\textit{c.f.,} Figure~\ref{t_ch}~top row).

For the other models, $t^\star$ is determined by the complex combination of many different competing mechanisms. For example, Figure~\ref{qvsdp_tch} shows that the inclusion of Kozeny-Carman permeability increases $t^\star$ relative to the corresponding constant-permeability models for the fixed $\Delta{p}$ cases, but that the opposite is true for the fixed $q$ cases (\textit{i.e.}, the -$k_\mathrm{KC}$ models evolve slower than the corresponding -$k_0$ models for fixed $\Delta{p}$, but faster for fixed $q$).

In general, we expect an increase or decrease in the effective permeability of the medium to lead to a faster or slower evolution (a decrease or increase in $t^\star$) respectively, because the characteristic poroelastic timescale is inversely proportional to permeability, $\tilde{T}_\mathrm{pe}\propto{}1/\tilde{k}$ (\textit{c.f.,} \S\ref{scaling}). However, the permeability field is transient and non-uniform. The development of localised regions with significantly reduced permeability can act as a global bottleneck because of the harmonic nature of combining permeabilities in series, leading to a slower evolution (increase in $t^\star$) even when the permeability increases in most of the domain. In the fixed $q$ case, the effective permeability of the medium increases monotonically in time and therefore the impact of deformation-dependent permeability is to speed up the evolution (decrease $t^\star$). In the fixed $\Delta{p}$ case, in contrast, the sudden development of a sharp, low-porosity boundary layer at the outer boundary dominates the effective permeability of the medium, leading to a strong transient decrease, and this slows the evolution (increases $t^\star$). This effect is particularly strong because the boundary layer significantly overshoots the steady-state porosity.

Nonlinear kinematics enter the problem in several ways competing ways. One major difference between the L models (linearised kinematics) and the Q and N models (rigorous kinematics) is the relationship between $\partial{u_s}/\partial{t}$ and $v_s$. The linearised relationship is $\partial{u_s}/\partial{t}=v_s$, whereas the exact relationship is $\partial{u_s}/\partial{t}=(1/\lambda_r)v_s$. Since $v_s=q/r+k(\phi_f)(\partial{p}/\partial{r})$ in all models, the additional factor of $1/\lambda_r=1-\partial{u_s}/\partial{r}$ typically accelerates the deformation in the Q and N models relative to the L models because, in most cases, $\partial{u_s}/\partial{r}<0$ (\textit{c.f.,} Figures \ref{all_quants_dp_vs_q} and \ref{all_quants_apvszd}), and therefore $1/\lambda_r>1$. That is, linearisation leads to an underestimation of $\partial{u_s}/\partial{t}$ in the L models and we would therefore expect the introduction of nonlinear kinematics to speed up the evolution (decrease $t^\star$). A second effect is that the inner boundary condition is applied at the moving inner boundary in the Q and N models, but at the original position of the inner boundary in the L models. For the fixed $\Delta{p}$ case, the outward motion of the inner boundary should increase $q$ (applying the same $\Delta{p}$ across a thinner wall leads to a larger $q$), which would imply a faster evolution (smaller $t^\star$) in the Q and N models relative to the L models. For the fixed $q$ case, the outward motion of the inner boundary should lead to a lower $\Delta{p}$ across the material (driving the same $q$ through a thinner wall requires a smaller $\Delta{p}$), which would imply a slower evolution (larger $t^\star$) in the Q and N models relative to the L models. These latter effects, although clearly weak, are visible in Figure~\ref{qvsdp}. A third effect is simply that the Q and N models deform more than the L models for this scenario~\citep{auton2017arteries}; as a result, it takes longer for the Q and N models to reach a certain relative difference from their respective steady states (larger $t^\star$). These three effects combine such that nonlinear kinematics speed up the deformation for fixed $\Delta{p}$ (the first effect dominates), but slow down the deformation for fixed $q$ (the second and third effects dominate) (Figure~\ref{qvsdp_tch}).

Finally, we consider the role of nonlinear elasticity. The L and Q models use linear elasticity, whereas the N models uses Hencky elasticity. The elasticity law determines the relationship between $u_s$ and $\partial{p}/\partial{r}$, and also plays a role in the boundary conditions. Under uniform uniaxial deformation, Hencky elasticity is stiffer than linear elasticity in compression and softer than linear elasticity in tension (\textit{c.f.,} Figure~A1 of the electronic supplementary material (ESM) of \citet{auton2017arteries}). The classical poroelastic timescale is inversely proportional to stiffness, $\tilde{T}_\mathrm{pe}\propto{}1/\tilde{\mathcal{M}}$ (\textit{c.f.,} \S\ref{scaling}), so we would expect the N models to deform more quickly than the L and Q models in uniaxial scenarios dominated by compression, and more slowly than the L and Q models in uniaxial scenarios dominated by tension. It is not straightforward to extrapolate these expectations to a biaxial problem featuring a mix of tension and compression, but we note that the N models does evolve slightly more quickly than the Q models in the fixed $\Delta{p}$ case, which features strong transient compression in $\sigma_r^\prime$ near both boundaries; this is consistent with the general explanation above.

\subsection{Impact of constraint on thin-walled cylinders for fixed $q$}

\begin{figure}[tb]
    \centering
    \includegraphics[width=5in]{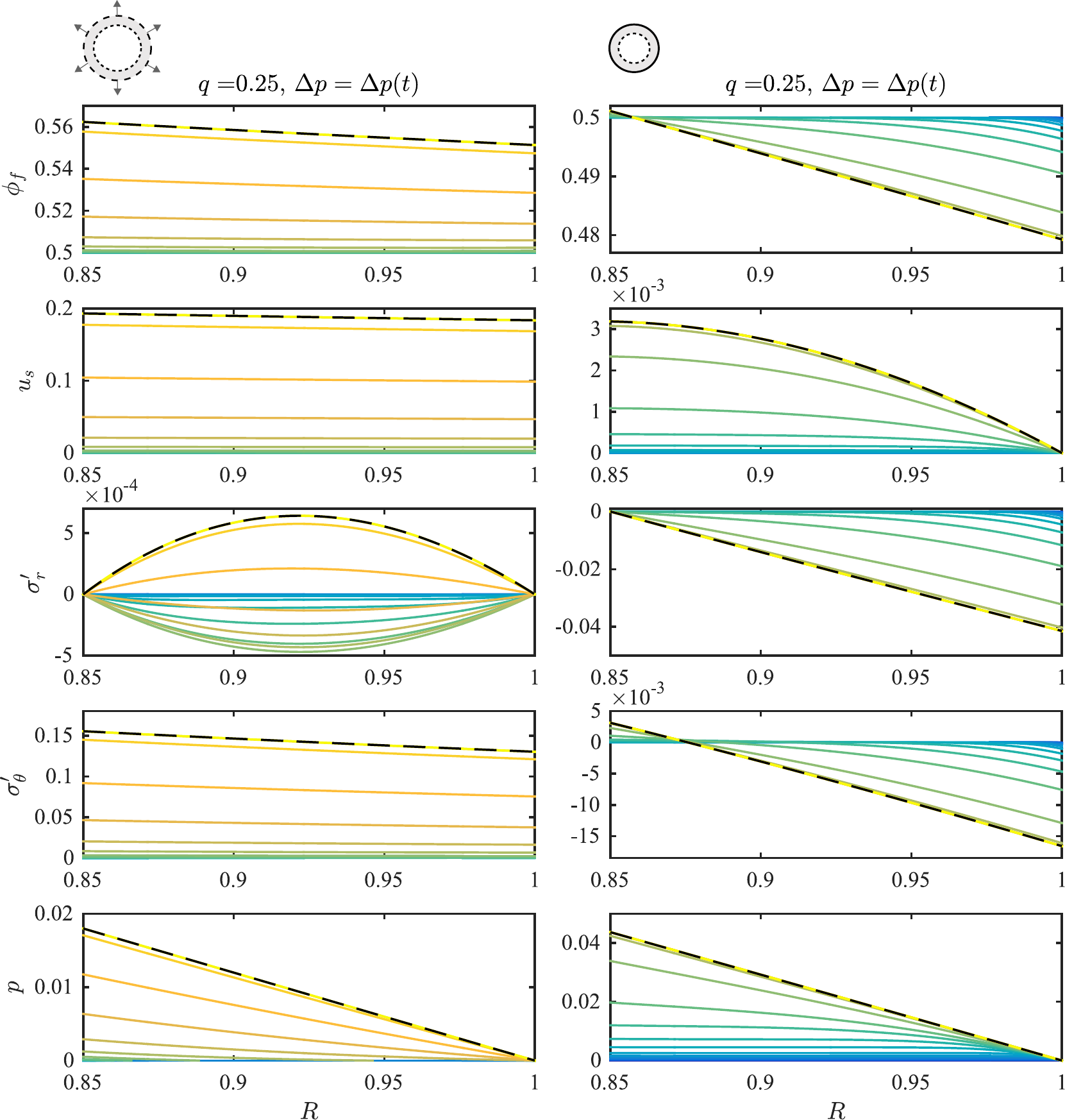}
    \caption{The time evolution of an unconstrained thin-walled cylinder (left column) and a constrained thin-walled cylinder (right column), for the Q-$k_\mathrm{KC}$ model ($a_0 = 0.85$), where the flow is driven by fixed $q=0.25$. We show the solution at a range of times, logarithmically spaced from $t=10^{-7}$ (blue) to $t=10$ (yellow). We also show the initial condition and the steady state for reference (solid and dashed black lines, respectively). The constrained cylinder reaches steady state much more quickly than the unconstrained cylinder. \label{all_quants_apvszd} }
\end{figure}

\begin{figure}[tb]
    \centering
    \includegraphics[width=5in]{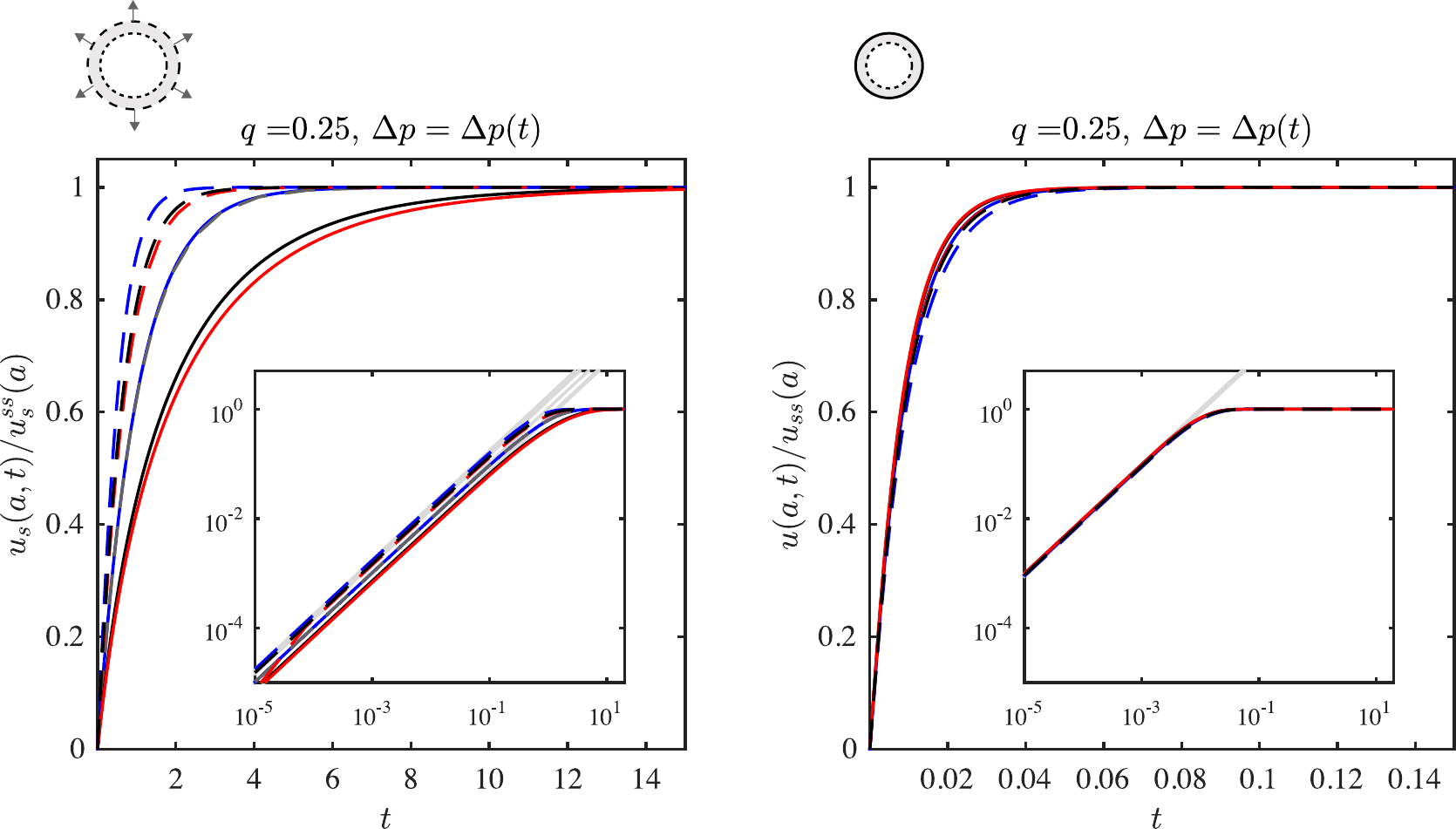}
    \caption{The time evolution of the normalised displacement at the inner boundary for all six models for all six models for the same scenario and parameter values as in Figure~\ref{all_quants_apvszd}. Line colours and styles are the same as in Figure~\ref{dpvsq_u_a}--\ref{qvsdp_tch}, although we additionally show the analytical dynamic L-$k_0$ solution (dashed grey lines). The insets show the evolution on a logarithmic scale, clearly demonstrating that $u_s(a,t)\sim{}(q/a_0)t$ for $t\ll{}1$ (solid grey lines) in all cases. Note that the time evolution of the unconstrained cylinder is about two orders of magnitude slower than that of the constrained cylinder, and that model choice is much more important for the unconstrained cylinder than for the constrained cylinder for these parameters. \label{quantify} }
\end{figure}

\begin{figure}[tb]
    \centering
    \includegraphics[width=5in]{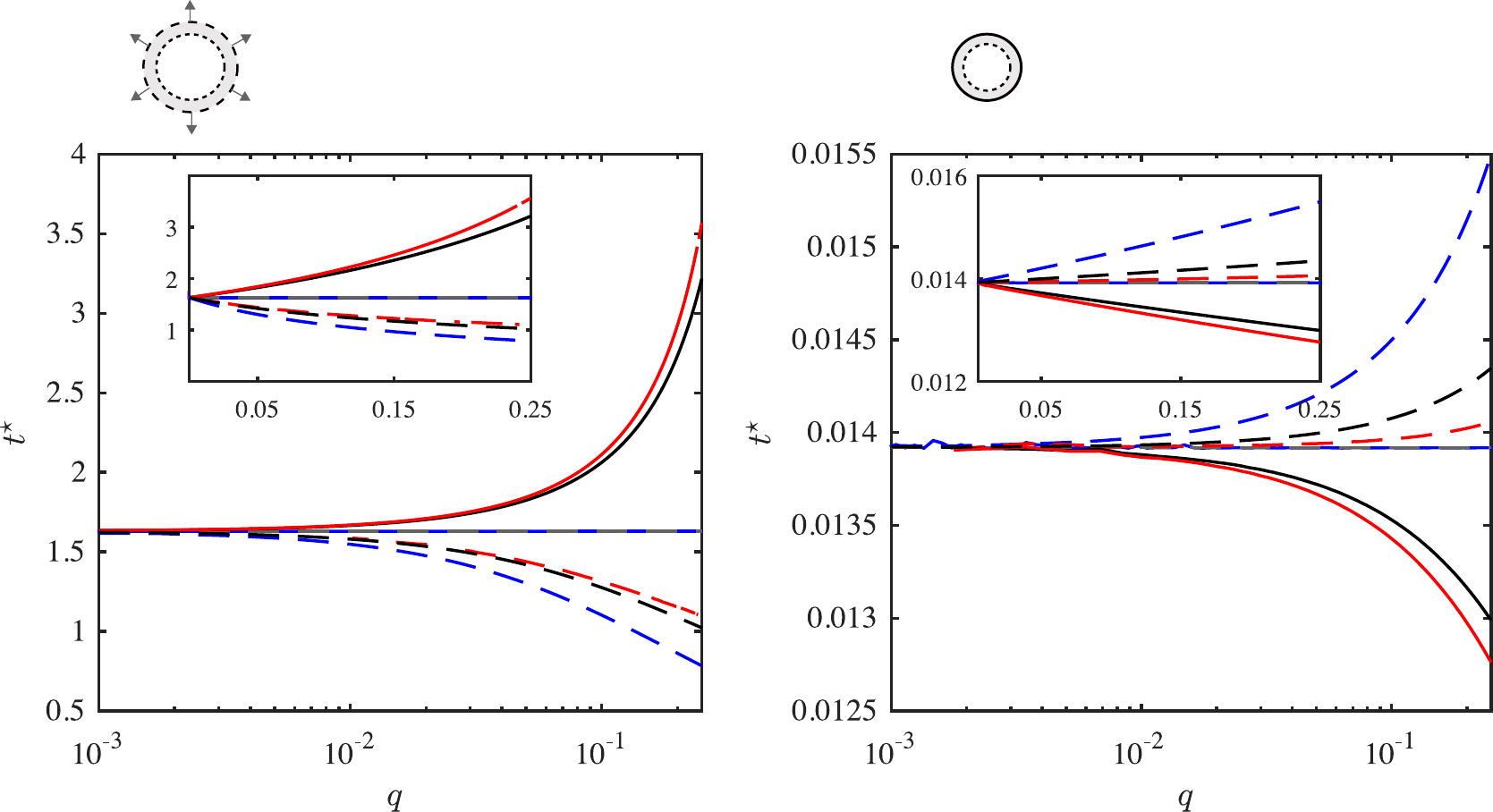}
    \caption{The evolution timescale $t^\star$ of the thin-walled cylinder ($a_0=0.85$) for all six models for unconstrained (left) and constrained (right) cases. Line colours and styles are the same as in Figures \ref{dpvsq_u_a}--\ref{qvsdp_tch} and \ref{quantify}. The evolution timescale of the L-$k_0$ model is again independent of driving strength (\textit{c.f.}, Figure~\ref{qvsdp_tch}). For the unconstrained cylinder, $t^\star$ increases strongly with $q$ for the Q-$k_0$ and N-$k_0$ models and decreases strongly with $q$ for all of the -$k_\mathrm{KC}$ models; for the constrained cylinder, these effects are reversed in direction and much smaller in magnitude. To highlight the extent to which nonlinearity accelerates or decelerates the transient evolution in each case relative to the L-$k_0$ model, we plot these same results as $t^\star/t^\star_{\mathrm{L}k0}$ against $q$ and $\Delta{p}$ in Appendix~E (Figure~E2). \label{ap_vs_zd_tch} }
\end{figure}

We now consider the impact of the outer boundary condition on the transient evolution of the problem. We do so by comparing the behaviour of an unconstrained thin-walled cylinder to that of a constrained thin-walled cylinder ($a_0=0.85$) for fixed $q$. In Figure~\ref{all_quants_apvszd}, we show the time evolution of all key quantities for the Q-$k_\mathrm{KC}$ model. In Figure~\ref{quantify}, we examine the impact of model choice by considering the evolution of the displacement at the inner boundary for all six models. In Figure~\ref{ap_vs_zd_tch}, we consider the characteristic evolution timescale for this geometry by comparing these boundary conditions over a wide range of $q$ values for all six models.

In Figure~\ref{all_quants_apvszd}, we plot the evolution of all key quantities to steady state for the Q-$k_\mathrm{KC}$ model for fixed $q=0.25$ for an unconstrained thin-walled cylinder (left column) and for a constrained thin-walled cylinder (right column) ($a_0=0.85$). For the unconstrained cylinder, all quantities except for $\sigma_r^\prime$ evolve monotonically in time. The radial effective stress $\sigma_r^\prime$ initially decreases into compression throughout entire the interior of the cylinder, before eventually increasing to its tensile steady state. Note, however, that $||\sigma_r^\prime||$ is several orders of magnitude smaller than $||\sigma_\theta^\prime||$. For the constrained cylinder, all quantities except for $\sigma_\theta^\prime$ evolve monotonically in time, and $\sigma_\theta^\prime$ evolves monotonically except in a small region near the transition from tension to compression. More importantly, all quantities evolve much more quickly than for the unconstrained cylinder.

In Figure \ref{quantify}, we again consider the normalised displacement at the inner boundary $\bar{u}_a$ for all six models, comparing the evolution of $\bar{u}_a$ for the unconstrained (left) and constrained (right) cylinders. The evolution is monotonic for all models in both cases, but about two orders of magnitude faster for the constrained cylinder than for the unconstrained cylinder. Plotting $\bar{u}_a$ on a logarithmic scale (insets) highlights its early-time evolution, which is shown to be $u_s(a,t)\sim{}(q/a_0)t$ for $t\ll{}1$ for all $a_0$ (see Appendix~D). For the unconstrained cylinder, the Q-$k_0$ and N-$k_0$ models clearly evolve much more slowly than the L-$k_0$ model and all -$k_0$ models are much slower than the -$k_\mathrm{KC}$ models. Note also that the models are in the same relative order as in Figure~\ref{dpvsq_u_a}. For the constrained cylinder, in contrast, the rate is relatively insensitive to model choice and the ordering of the models is reversed.

In Figure~\ref{ap_vs_zd_tch}, we consider the characteristic evolution timescale $t^\star$ for a thin-walled cylinder for all six models as a function of $q$. For the unconstrained cylinder (left), $t^\star$ increases with $q$ for the Q-$k_0$ and N-$k_0$ models, but decreases with $q$ for the -$k_\mathrm{KC}$ models. This suggests that, for an unconstrained cylinder, nonlinear kinematics speed up the deformation whereas deformation-dependent permeability slows down the deformation. The underlying physical mechanisms are the same as those discussed in regard to Figure~\ref{qvsdp_tch}. The various impacts of these mechanisms are easier to interpret in this case because the permeability increases strongly throughout the material, $\partial{u_s}/\partial{r}$ is strictly negative, and the inner radius moves substantially. These various impacts are also much stronger in this case because the deformation is much larger, and they are further amplified as $q$ increases---$t^\star$ changes by several-fold over this range of $q$. The Q and N models are again very similar, suggesting that the elasticity law again plays a relatively minor role. For the constrained cylinder (right), all of these effects are reversed in direction and much smaller in magnitude, with $t^\star$ changing by up to about 10\% over this range of $q$. This is straightforward to interpret: the permeability decreases strongly almost everywhere, and the displacement is two orders of magnitude smaller than for the unconstrained cylinder. For the L-$k_0$ model, $t^\star$ is independent of $q$ for both unconstrained and constrained cylinders. Lastly, note that the constrained cylinder evolves about two orders of magnitude faster than the unconstrained cylinder, which is simply due to the fact that the unconstrained cylinder deforms substantially more in total.

\subsection{Impact of geometry and driving method on time evolution}

\begin{figure}[tb]
    \centering
    \includegraphics[width=5in]{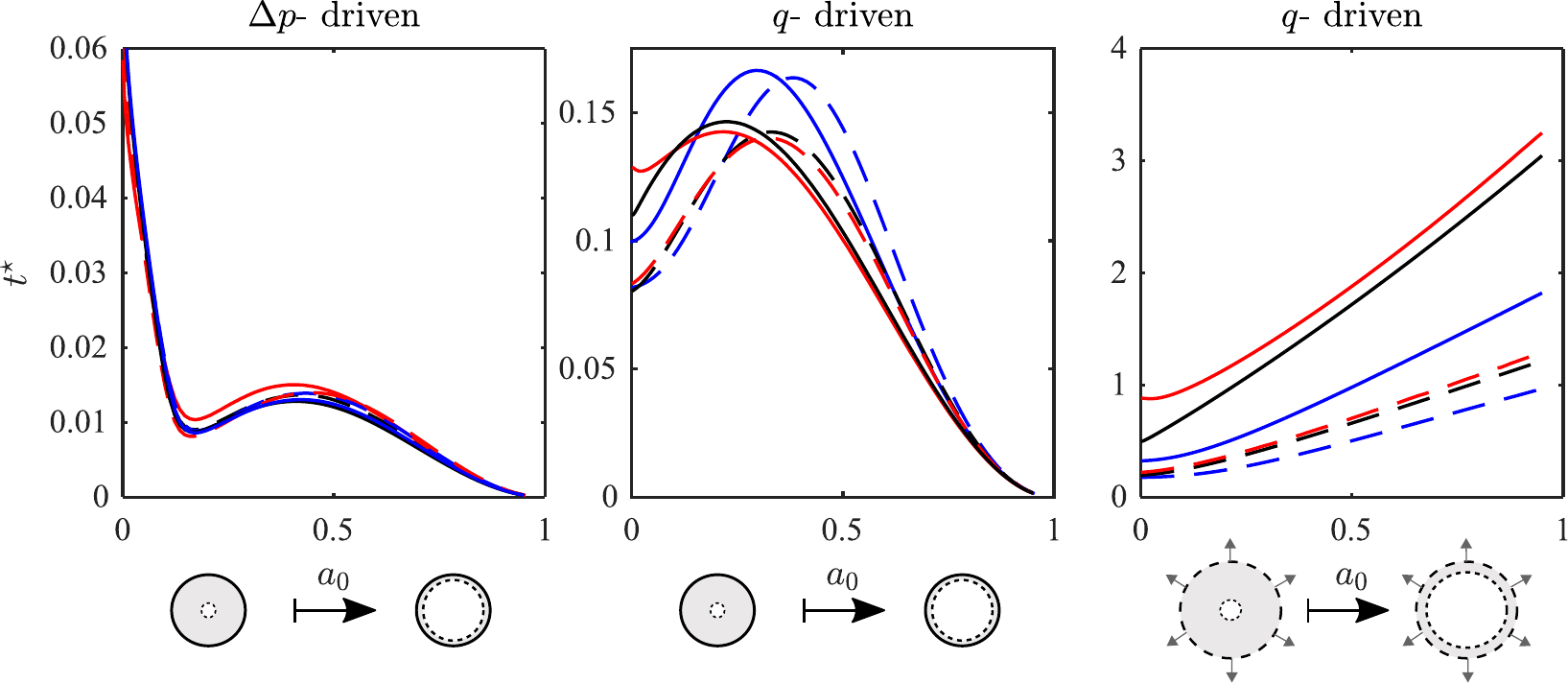}
    \caption{The impact of $a_0$ on $t^\star$ for a constrained cylinder driven by fixed $\Delta{p}=0.05$ (left), a constrained cylinder driven by fixed $q=0.2$ (centre), and an unconstrained cylinder driven by fixed $q=0.2$ (right), and for all 6 models in each case. Line colours and styles are the same as in Figures~\ref{dpvsq_u_a}--\ref{qvsdp_tch} and \ref{quantify}--\ref{ap_vs_zd_tch}. For the constrained cylinder, the evolution is fastest for thin walls; for the unconstrained cylinder, it is fastest for thick walls. Note also that $t^\star$ depends non-monotonically on $a_0$ for the constrained cylinder, and is particularly sensitive to model choice for fixed $q$. To highlight the extent to which nonlinearity accelerates or decelerates the transient evolution in each case relative to the L-$k_0$ model, we plot these same results as $t^\star/t^\star_{Lk0}$ against $q$ and $\Delta{p}$ in Appendix~E (Figure~E3). \label{t_ch} }
\end{figure}

In Figure~\ref{t_ch}, we plot the evolution time $t^\star$ against $a_0$ for all six models for constrained cylinders for fixed $\Delta{p}$ (left) and fixed $q$ (centre), and for unconstrained cylinders for fixed $q$. Note, firstly, that $t^\star$ is a strong function of $a_0$ even for classical linear poroelasticity (the L-$k_0$ model). This is evident from the time-dependent component of the analytical solution, which is a nontrivial function of $a_0$. For the constrained cylinder for fixed $\Delta{p}$, model choice appears to be relatively unimportant relative to geometry; the evolution is slowest for thick walls and fastest for thin walls, but $t^\star$ has a modest local maximum around $a_0\approx{}0.2$. For the constrained cylinder for fixed $q$, model choice is much more important; the evolution is still fastest for thin walls, but $t^\star$ now has a global maximum for some intermediate wall thickness and the amplitude and position of this maximum varies strongly from model to model. Importantly, the relative roles of deformation-dependent permeability, nonlinear kinematics, and nonlinear elasticity are nontrivial and depend strongly on $a_0$---this suggests that the conclusions drawn above for constrained cylinders are valid only for thick walls. For the unconstrained cylinder for fixed $q$, $t^\star$ increases monotonically with $a_0$ for all models except the N-$k_0$ model, which exhibits a weak interior minimum at $a_0\approx{}0.05$. The fast evolution occurs for thick-walled cylinders and the slowest for thin-walled cylinders, and the ordering of the models is independent of $a_0$. As a result, the conclusions drawn above for unconstrained cylinders should be qualitatively valid for all $a_0$. Finally, note that $t^\star$ is about an order of magnitude larger for constrained cylinders for fixed $q$ than for constrained cylinders for fixed $\Delta{p}$, and an about order of magnitude larger again for unconstrained cylinders for fixed $q$ (except when the walls are very thick, in which case the dynamics of constrained and unconstrained cylinders for fixed $q$ are very similar; see Appendix~F (Figure~F1)).

\section{Conclusion}

Despite being central to important problems in a variety of fields, radial poroelastic deformations remain relatively poorly understood. Here, we have presented the first systematic exploration of the effects of nonlinearity, driving method, and geometry on the transient evolution of these deformations. We have shown that the radial geometry results in nontrivial biaxial deformations and a strong dependence on wall thickness that render this problem much more complex and nuanced than the corresponding uniaxial scenario~\citep{macminn2016large}.

We first illustrated the strong qualitative and quantitative impacts of driving method (fixed $\Delta{p}$ \textit{vs.} fixed $q$). For a constrained cylinder with thick walls, we showed that driving with fixed $\Delta{p}$ leads to non-monotonicity in the time evolution of the displacement at the inner boundary for all models, even classical linear poroelasticity (Figure~\ref{dpvsq_u_a}). This does not occur when driving with fixed $q$. We considered the details of these deformations in the context of a model that includes rigorous nonlinear kinematics and deformation-dependent permeability, but with the simplification of linear elasticity (Q-$k_\mathrm{KC}$) (Figure~\ref{all_quants_dp_vs_q}). We found that the nonmonotonicity mentioned above was reflected in the azimuthal effective stress, again even for linear poroelasticity, with implications for applications such as hydraulic fracturing. We found that this scenario evolves much more quickly when driven by fixed $\Delta{p}$ as opposed to fixed $q$, and that this is true for all models across a wide range of driving values (Figure \ref{qvsdp_tch}). We also found that, when $\Delta{p}$ is fixed, $q(t)$ is initially very high and decreases towards the steady-state value; when $q$ is fixed, in contrast, $\Delta{p}(t)$ is initially very small and increases towards the steady-state value (Figure \ref{qvsdp}).

We also investigated the impact of constraint at the outer boundary (unconstrained \textit{vs.} constrained). For a thin-walled cylinder, we found that a constrained cylinder evolves about two orders of magnitude more quickly than an unconstrained cylinder (Figures~\ref{all_quants_apvszd}--\ref{ap_vs_zd_tch}). We also found that the evolution timescale of an unconstrained thin-walled cylinder is dominated by kinematics and deformation-dependent permeability (Figures \ref{quantify}- \ref{ap_vs_zd_tch}), despite the fact that the steady state is dominated by kinematics and elasticity \citep{auton2017arteries}.

Finally, we showed that the evolution timescale depends strongly on wall thickness for all models, for both driving conditions, and for both constrained and unconstrained cylinders (Figure~\ref{t_ch}). For constrained cylinders, the evolution timescale is nonmonotic in wall thickness for all models and for both driving conditions. For unconstrained cylinders driven by fixed $q$, the relative contributions of nonlinear kinematics, deformation-dependent permeability, and nonlinear elasticity also depend very strongly on wall thickness.



The authors are grateful to EPSRC for support in the form of a Doctoral Training Award to LCA. The authors also thank Ian Griffiths and Andrew Wells for helpful discussions.

%

\clearpage
\appendix

\renewcommand{\theequation}{\thesection\arabic{equation}}
\renewcommand{\thefigure}{\thesection\arabic{figure}}
\setcounter{figure}{0}

\section{Analytical solution: L-$k_0$ for fixed $q$ and $\sigma_r^\prime(b,t)=\sigma_r^\star$}\label{lk0apana}

Here we present the analytical solution for a cylinder subjected to the linearised applied-stress conditions $B^{a_0}[u_s]=B^{b_0}_2[u_s]=0$ for the general case $\sigma^\star \not\equiv0$. A stress-free initial condition does not satisfy these boundary conditions, so we begin by solving the corresponding non-poroelastic (``drained'') problem subjected to $B^{b_0}_2[u_s]=0$. We then use this solution as a consistent initial condition for the evolution of the fluid-driven problem. For the L-$k_0$ model, this initial condition corresponds to a uniform porosity that is distinct from the reference porosity $\phi_{f,0}$. Equation~(3.1a) subjected $B^{a_0}[u_s] = B^{b_0}_2[u_s] = 0$ leads to the IBVP \begin{subequations}
\begin{align}
    \label{TLb}
\frac{\partial{u_s}}{\partial{t}} - \frac{\partial }{\partial{r}}\left[\frac{1}{r}\frac{\partial}{\partial{r}}(u_sr)\right] = \frac{q}{r}&\qquad t>0\quad a_0< r< 1, \\
\label{ICb}
{u_s(r,0) = u_{s,0}= \frac{\sigma_r^\star }{1-a_0^2}\left[\frac{r}{1+\Gamma}+\frac{a_0^2}{(1-\Gamma)r}\right]} & {\qquad t=0\quad a_0< r< 1}, \\
\label{BC1b}
B^{a_0}[u_s] =\frac{\partial{u_s}}{\partial{r}}+ \Gamma\frac{u_s}{a_0}= 0 &\qquad t>0\quad r=a_0, \\
\label{BC2b}
B^{b_0}_2[u_s] = \frac{\partial{u_s}}{\partial{r}}+ \Gamma{u_s}- \sigma_r^\star = 0 &\qquad t>0\quad r=1.
\end{align}
\end{subequations}
Unlike for the constrained cylinder, the outer boundary condition in the above problem is not homogeneous. To address this, we begin by decomposing $u_s(r,t)$ into two functions $v(r,t)$ and $w(r,t)$, such that $u_s(r,t)=v(r,t)+w(r,t)$ and where $w(r,t)$ is chosen to satisfy the non-homogeneous boundary condition (Eq.~\ref{BC2b}),
\begin{equation}
    w(r,t) = \frac{\sigma_r^\star}{a_0^2-1}\left(\frac{a_0^2}{\Gamma}-\frac{r^2}{2+\Gamma}\right).
\end{equation}
Hence, $v(r,t)$ must then satisfy the IBVP
\begin{subequations}
\begin{align}
\label{TLc}
\frac{\partial v}{\partial{t}} - \frac{\partial }{\partial{r}}\left[\frac{1}{r}\frac{\partial}{\partial{r}}(vr)\right] = \frac{q}{r}- \frac{\sigma_r^\star}{a_0^2-1}\left(\frac{3}{2+\Gamma}+\frac{a_0^2}{r^2\Gamma}\right)& \qquad t>0\quad a_0\leq r\leq 1, \\
\label{ICc}
v(r,0)= \frac{\sigma_r^\star}{1-a_0^2}\left[\frac{r}{1+\Gamma}+\frac{a_0^2}{(1-\Gamma)r}+\frac{a_0^2}{\Gamma}-\frac{r^2}{2+\Gamma}\right]& \qquad t=0\quad a_0\leq r\leq 1,
\\
\label{BC1c}
\frac{\partial{v}}{\partial{r}}+\Gamma\frac{v}{a_0}=0 & \qquad t>0\quad r=a_0,
\\
\label{BC2c}
\frac{\partial{v}}{\partial{r}}+\Gamma{v}=0& \qquad t>0\quad r=1,
\end{align}
\end{subequations}
which comprises a non-homogeneous PDE with a spatially dependent initial condition and, importantly, two homogenous boundary conditions (both Robin). We can now solve the problem as in \S3(b) via separation of variables. We begin with the separable ansatz
\begin{equation}\label{ansatz}
    v = \mathcal{T}(t)\mathcal{R}(r)
\end{equation}
on the associated homogeneous PDE
\begin{equation}\label{homo}
    \frac{\partial v}{\partial{t}} - \frac{1}{r} \left[\frac{\partial }{\partial{r}}\left(r\frac{\partial v}{\partial{r}}\right) -\frac{v}{r}\right] = 0.
\end{equation}
This allows us to consider the spatial problem as a Sturm-Liouville eigenvalue problem,
\begin{subequations}
\begin{equation}
    \mathcal{R}''(r)+\frac{\mathcal{R}'(r)}{r} +\mathcal{R}(r)\left(\omega^2-\frac{1}{r^2}\right) = 0,
\end{equation}
\begin{equation}\label{BCSc}
    \mathcal{R}'(a_0)+\frac{\Gamma}{a_0}\mathcal{R}(a_0)= \mathcal{R}'(1)+\Gamma \mathcal{R}(1)=0,
\end{equation}
\end{subequations}
for some constant $\omega$. From this, the weighting function for the associated orthogonality condition is, once again, $r$. We solve for $\mathcal{R}(r)$ to obtain 
\begin{equation}
\mathcal{R}(r)=c_1J_1(\omega{}r)+c_2Y_1(\omega{}r),
\end{equation}
where $c_1$ and $c_2$ are constants to be determined. Using Equation~(\ref{BCSc}), we obtain 
\begin{equation}\label{eq:2x2aac}
    \left(\begin{array}{cc}
        \mathcal{A} & \mathcal{B} \\
        \mathcal{C} & \mathcal{D}
    \end{array}\right)
    \left(\begin{array}{c}
        c_1 \\
        c_2
    \end{array} \right) = 
    \left(\begin{array}{c}
        0 \\
        0
    \end{array} \right)
\end{equation}
where
\begin{subequations}\label{eq:ABCD}
    \begin{align}
        \mathcal{A}& \defeq \frac{\omega{}}{2}[J_0(\omega{}a_0)-J_2(\omega{}a_0)]+\frac{\Gamma}{a_0}J_1(\omega{}a_0), \\
        \mathcal{B}& \defeq  \frac{\omega{}}{2}[Y_0(\omega{}a_0)-Y_2(\omega{}a_0)]+\frac{\Gamma}{a_0}Y_1(\omega{}a_0), \\
        \mathcal{C}& \defeq \frac{\omega{}}{2}[J_0(\omega{})-J_2(\omega{})]+\Gamma J_1(\omega{}), \\
        \mathcal{D}& \defeq \frac{\omega{}}{2}[Y_0(\omega{})-Y_2(\omega{})]+\Gamma Y_1(\omega{}).
    \end{align}
\end{subequations}
In order for Equation~\eqref{eq:2x2aac} to have a non-trivial solution, it must be the case that $\mathcal{A}\mathcal{D}-\mathcal{B}\mathcal{C} = 0$,
\begin{multline}\label{detc}
\left\{\frac{\omega{}}{2}\left[J_0(\omega{}a_0)-J_2(\omega{}a_0)\right]+\frac{\Gamma}{a_0}J_1(\omega{}a_0)\right\}\left\{\frac{\omega{}}{2}\left[Y_0(\omega{})-Y_2(\omega{}b_0)\right]+\Gamma Y_1(\omega{})\right\}-\\
\left\{\frac{\omega{}}{2}[Y_0(\omega{}a_0)-Y_2(\omega{}a_0)]+\frac{\Gamma}{a_0}Y_1(\omega{}a_0)\right\}\left\{\frac{\omega{}}{2}\left[J_0(\omega{})-J_2(\omega{})\right]+\Gamma J_1(\omega{})\right\} = 0.
\end{multline}
Equation~\eqref{detc} suggests an infinite number of solutions for an infinite number of distinct eigenvalues $\omega = \omega_n$. The boundary condition at $r=a_0$ (Equation~\ref{BC1c}) yields
\begin{equation}
    c_2 = -c_1\frac{\omega{}a_0[J_0(\omega{}a_0)-J_2(\omega{}a_0)]+2\Gamma J_1(\omega{}a_0)}{\omega{}a_0[Y_0(\omega{}a_0)-Y_2(\omega{}a_0)]+2\Gamma Y_1(\omega{}a_0)} =-c_1\frac{\mathcal{A}}{\mathcal{B}},
\end{equation}
giving the associated infinite number of eigenfunctions $\mathcal{R}_n(r)$ as
\begin{equation}
\mathcal{R}_n(r) = J_1(\omega_nr)-\frac{\mathcal{A}}{\mathcal{B}} Y_1(\omega_nr).
\end{equation}
The general series solution for $v(r,t)$ is then
\begin{equation}
    v(r,t) = \sum_{n=1}^{\infty}\mathcal{T}_n(t)\mathcal{R}_n(r),
\end{equation}
where $\mathcal{T}_n(t)$ must be determined to satisfy Equation~(\ref{TLc}). This requirement leads to
\begin{equation}\label{eq:requ}
    \frac{q}{r}- \frac{\sigma_r^\star}{a_0^2-1}\left(\frac{3}{2+\Gamma}+\frac{a_0^2}{r^2\Gamma}\right) = \frac{\partial v}{\partial{t}} - \frac{\partial }{\partial{r}}\left[\frac{1}{r}\frac{\partial}{\partial{r}}(vr)\right] =\sum_{n=1}^{\infty}\left[\mathcal{T}'_n(t)+\omega_n^2\mathcal{T}_n(t)\right] \mathcal{R}_n(r),
\end{equation}
motivating the decomposition
\begin{equation}\label{eq:decomp}
    \frac{q}{r}- \frac{\sigma_r^\star}{a_0^2-1}\left(\frac{3}{2+\Gamma}+\frac{a_0^2}{r^2\Gamma}\right) = \sum_{n=1}^{\infty} f_n \mathcal{R}_n(r)
\end{equation}
for some infinite set of constants $f_n$. Equations~\eqref{eq:requ} and \eqref{eq:decomp} lead to
\begin{equation}
    \mathcal{T}'_n(t)+\omega_n^2\mathcal{T}_n(t)=f_n,
\end{equation}
subject to $\mathcal{T}_n(0) = t_n$, where $t_n$ is an infinite set of constants to be determined. This has solution
\begin{equation}
    {\mathcal{T}_n(t)=\frac{f_n}{\omega_n^2}\left(1-e^{-\omega_n^2t}\right) + t_ne^{-{\omega_n^2t}}},
\end{equation}
where, due to the properties developed in Equations~(3.15), $t_n$ is given by

\begin{equation}\label{notsosimples}
    t_n =\frac{\displaystyle\int_{a_0}^1rv(r,0)\mathcal{R}_n\mathrm{d}r}{\displaystyle\int_{a_0}^1[\mathcal{R}_n(r)]^2r\mathrm{d}r}= \frac{\sigma_r^\star }{1-a_0^2}\left(\frac{\displaystyle\frac{1}{1+\Gamma}I_7 +\displaystyle\frac{a_0^2}{1-\Gamma}I_5-\displaystyle\frac{1}{2+\Gamma}I_8+\displaystyle\frac{a_0^2}{\Gamma}I_6}{I_1 +\displaystyle\frac{\mathcal{A}^2}{\mathcal{B}^2}I_2 - \displaystyle\frac{2\mathcal{A}}{\mathcal{B}}I_3}\right),
\end{equation}
where $I_1$, $I_2$, and $I_3$ are defined in Eqs.~(3.22) and
\begin{subequations}
\label{Ints2}
\begin{align}
\begin{split}
I_4 &\defeq \int_{a_0}^1\frac{\mathcal{R}_n(r)}{r}\mathrm{d}r = \left[\omega_nr\left\{J_0(\omega_nr)-\frac{\mathcal{A}}{\mathcal{B}}Y_0(\omega_nr)\right\}- J_1(\omega_nr)+\frac{\mathcal{A}}{\mathcal{B}}Y_1(\omega_nr)\right]^1_{a_0} \\ & \hspace{8.8cm}+ {\omega_n} \int_{a_0}^1r\mathcal{R}_n(r)\mathrm{d}r,
\end{split} \\
\begin{split}
I_5 &\defeq \int_{a_0}^1\mathcal{R}_n(r)\mathrm{d}r = \frac{\Big[\mathcal{A}Y_0(\omega_nr)-\mathcal{B}J_0(\omega_nr)\Big]^1_{a_0}}{\omega_n\mathcal{B}},
\end{split} \\
\begin{split}
I_6 &\defeq \int_{a_0}^1r\mathcal{R}_n(r)\mathrm{d}r = \left[\frac{\pi r}{2\omega_n}\left\{\left[J_1(\omega_nr)-\frac{\mathcal{A}}{\mathcal{B}}Y_1(\omega_nr)\right]\boldsymbol{\mathrm{H}}_0(\omega_nr)\right.\right.\\ & \hspace{5.7cm}-\left.\left.\left[J_0(\omega_nr)-\frac{\mathcal{A}}{\mathcal{B}}Y_0(\omega_nr)\right]\boldsymbol{\mathrm{H}}_1(\omega_nr)\right\}\right]^1_{a_0}, 
\end{split}\\
\begin{split}
I_7 &\defeq \int_{a_0}^1r^2\mathcal{R}_n(r)\mathrm{d}r = \left[ \frac{r^2}{\omega_n} \left\{ J_2(\omega_nr)-\frac{\mathcal{A}}{\mathcal{B}}Y_2(\omega_nr)\right\}\right]_{a_0}^1,
\end{split}\\
\begin{split}
\nonumber
&\hspace{-.65cm}\text{and} 
\end{split}\\
\begin{split}
I_8 &\defeq \int_{a_0}^1r^3\mathcal{R}_n(r)\mathrm{d}r = \left[ \frac{3r^2}{{\omega_n}}\left\{ J_1(\omega_nr)-\frac{\mathcal{A}}{\mathcal{B}}Y_1(\omega_nr) \right\} \right. \\ & \hspace{3.8cm}-\left. \frac{r^3} {{\omega_n}} \left\{J_0(\omega_nr)-\frac{\mathcal{A}}{\mathcal{B}}Y_0(\omega_nr)\right\}\right]_{a_0}^1 -\frac{3}{{\omega_n}}\int_{a_0}^1r\mathcal{R}_n(r)\mathrm{d}r,
\end{split}
\end{align}
\end{subequations}
where $\boldsymbol{\mathrm{H}}_\nu(r)$ is the Struve function of order~$\nu$. Similarly,
\begin{equation}
\label{simplish}
f_n 
= \frac{qI_5+\displaystyle\frac{\sigma_r^\star}{1-a_0^2}\left( \displaystyle\frac{3}{2+\Gamma}I_6+\displaystyle\frac{a_0^2}{\Gamma}I_4 \right)}{I_1 +\displaystyle\frac{\mathcal{A}^2}{\mathcal{B}^2}I_2 - \displaystyle\frac{2\mathcal{A}}{\mathcal{B}}I_3}. 
\end{equation} 
The analytical solution is then finally given by
\begin{align}\label{anasolapstress}
    \begin{split}
        u_s(r,t) &= w(r,t)+v(r,t) \\
        &= \frac{\sigma_r^\star }{a_0^2-1}\left(\frac{a_0^2}{\Gamma}-\frac{r^2}{2+\Gamma}\right)
 + \sum_{n=1}^{\infty}
\left\{\left[\frac{f_n}{\omega_n}\left(1-e^{-\omega_nt}\right) - t_n{e^{-\omega_nt}} \right] 
\left[J_1(\omega_nr)-\frac{\mathcal{A}}{\mathcal{B}} Y_1(\omega_nr)\right]\right\},
    \end{split}
\end{align}
where the $\omega_n$ are the solutions to Equation~(\ref{detc}), $\mathcal{A}$ and $\mathcal{B}$ are defined in Equations~\eqref{eq:ABCD}, and the $t_n$ and the $f_n$ are defined by Equations~(\ref{notsosimples}) and (\ref{simplish}), respectively.

\section{Numerical implementation: $M$, $F$, and $c$ for fixed $q$}\label{mass}

Here we list the $F[u_s]$, $M[u_s]$ and $c[u_s]$ for all models for fixed $q$. Note that, although it is often preferable to simplify the mass matrix as much as possible, handling of the nonlinearity of $k[u_s]$ sometimes benefits from a more complicated mass matrix. We use the simplest possible mass matrices for all fixed $q$ cases except for applied-stress Q-$k_\mathrm{KC}$, applied-stress N-$k_\mathrm{KC}$, and constrained N-$k_\mathrm{KC}$. These three cases benefit from including $k[u_s]$ as a component in $M$ and $c$ rather than ${F}$.

\begin{subequations}
For the L models, we have
\begin{equation}
    M = 1,\quad F = k[u_s]\frac{\partial }{\partial{r}}\left[\frac{1}{r}\frac{\partial}{\partial{r}}(u_sr)\right] \quad \text{and} \quad c = 1.
\end{equation}

For the Q-$k_0$ models, we have
\begin{equation}
    M = 1,\quad F =\left(1-\frac{\partial{u_s}}{\partial{r}}\right)\left( \frac{\partial^2 u_s}{\partial{r^2}}+\frac{1}{r}\frac{\partial{u_s}}{\partial{r}}-\frac{u_s}{r^2}\right) \quad \text{and} \quad c = 1-\frac{\partial{u_s}}{\partial{r}}.
\end{equation}

For the Q-$k_\mathrm{KC}$ model subjected to $B^a[u_s] = B^b_1[u_s] =0$, we have 
\begin{equation}
 M = 1,\quad F =k[u_s]\left(1-\frac{\partial{u_s}}{\partial{r}}\right)\left( \frac{\partial^2 u_s}{\partial{r^2}}+\frac{1}{r}\frac{\partial{u_s}}{\partial{r}}-\frac{u_s}{r^2}\right) \quad \text{and} \quad c = 1-\frac{\partial{u_s}}{\partial{r}}.
\end{equation}

For the Q-$k_\mathrm{KC}$ model subjected to $B^a[u_s] = B^b_2[u_s] =0$, we have
\begin{equation}
    M = \frac{1}{k[u_s]},\quad F =\left(1-\frac{\partial{u_s}}{\partial{r}}\right)\left( \frac{\partial^2 u_s}{\partial{r^2}}+\frac{1}{r}\frac{\partial{u_s}}{\partial{r}}-\frac{u_s}{r^2}\right) \quad \text{and} \quad c = \frac{1-\displaystyle\frac{\partial{u_s}}{\partial{r}}}{k[u_s]}.
\end{equation}

For the N models, we first write the PDE explicitly in terms of $u_s$ in the following form:
\begin{subequations}
    \begin{align}
    \frac{\partial{u_s}}{\partial{t}} &=\frac{k[u_s]\left[1-\ln\left(\lambda_r\lambda_\theta^\Gamma\right)\right]}{\lambda_r\lambda_\theta}\left(\frac{\partial^2{u_s}}{\partial{r}^2} + \frac{\displaystyle\frac{q\lambda_r\lambda_\theta}{k[u_s]}-T}{B}\right), \\
    &T\defeq{}\left[\ln\left(\lambda_r\lambda_\theta^\Gamma\right) - \Gamma\right]\left(1-\frac{\lambda_\theta}{\lambda_r}\right) + (1-\Gamma)\ln\left(\frac{\lambda_\theta}{\lambda_r}\right), \\
    &B\defeq{} \lambda_r r\left[1-\ln\left(\lambda_r\lambda_\theta^\Gamma\right)\right].
    \end{align}
\end{subequations}

For the N-$k_0$ model, we have
\begin{equation}
    M = 1,\quad F = \frac{\left[1-\ln\left(\lambda_r\lambda_\theta^\Gamma\right)\right]}{\lambda_r\lambda_\theta} \left(\frac{\partial ^2 u_s}{\partial{r}^2} - \frac{T}{B} \right) \quad \text{and} \quad c = \frac{1}{\lambda_r r},
\end{equation}
    and, finally, for the N-$k_\mathrm{KC}$ model we have
\begin{equation}
    M = \frac{\lambda_r\lambda_\theta}{k[u_s]\left[1-\ln\left(\lambda_r\lambda_\theta^\Gamma\right)\right]}, \quad F = \frac{\partial ^2 u_s}{\partial{r}^2} - \frac{T}{B} \quad \text{and}\quad c = \frac{\lambda_r\lambda_\theta}{k[u_s]B}.
\end{equation}
\end{subequations}

\section{The evolution of $\sigma_\theta^\prime(a,t)$ for a constrained cylinder}
\begin{figure}[h]
    \centering
    \includegraphics[width=5in]{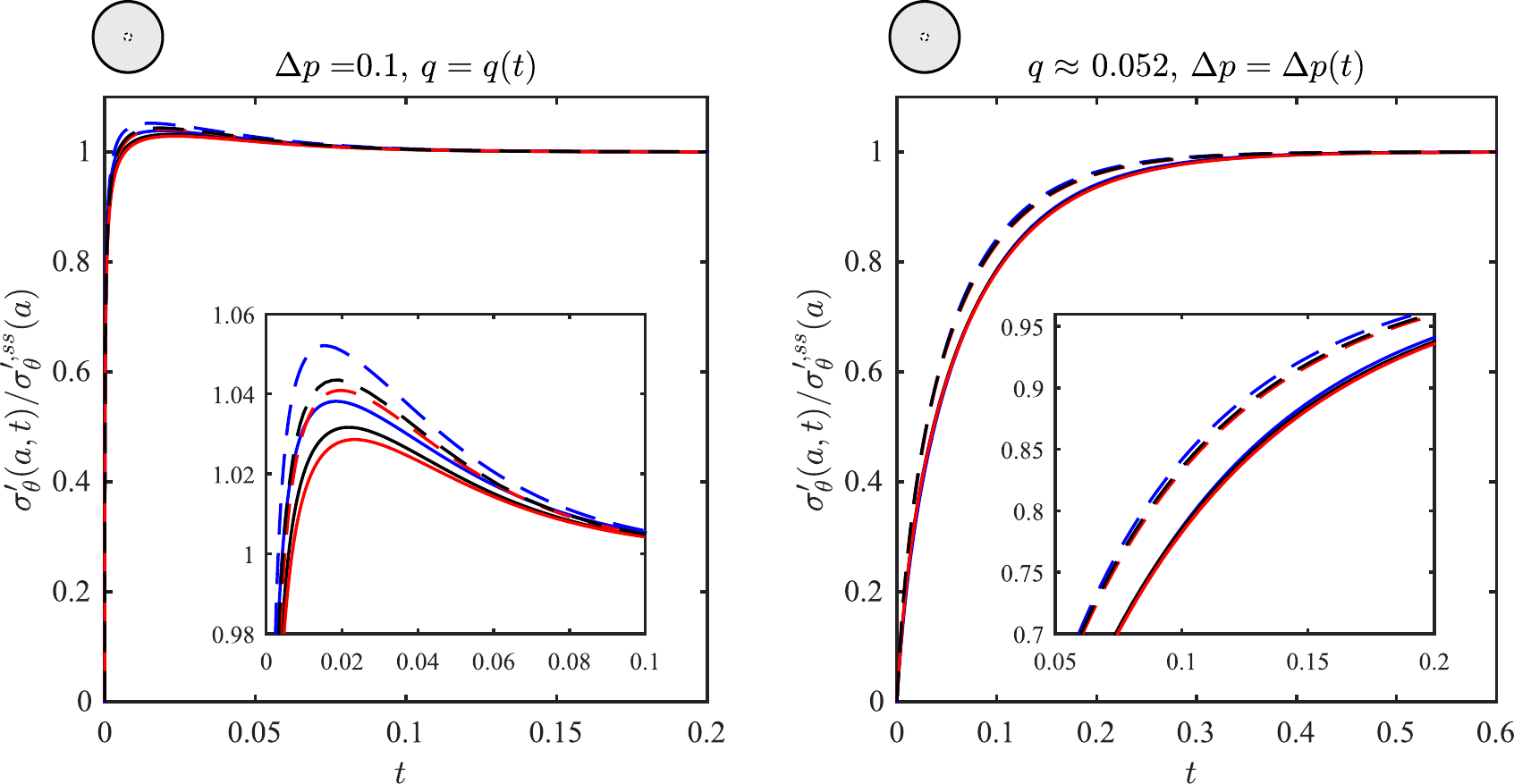}
    \caption{As Figure~3 of the main text, but for azimuthal effective stress rather than displacement. \label{dpvsq_u_a_supp} }
\end{figure}

In Figure~\ref{dpvsq_u_a_supp}, we present the transient evolution of the azimuthal effective stress at the inner boundary for a constrained cylinder (as Figure~3 of the main text, but for $\sigma_\theta^\prime$ rather than $u_s$). Note that $\sigma_\theta^\prime(a,t)$ exhibits the same behaviour as $u_s(a,t)$ for these parameters, in the sense that both are non-monotonic in time for fixed $\Delta{p}$ for all models, with a maximum at some intermediate time, and both are monotonic in time for fixed $q$.

\clearpage

\section{Early-time evolution of $u_s(a,t)$}

In Figure~7, we show that $u_s(a,t)\sim{}(q/a_0)t$ at early times for fixed $q$. We can derive this result for the L-$k_0$ model by considering Equation~(3.23) at early times. In general, the displacement at the inner boundary is given by
\begin{equation}
    u_s(t,a_0) = \sum_{n=1}^{\infty} \left[\frac{f_n}{\omega_n^2}\left(1-e^{-\omega_n^2{}t}\right)\right] \left[J_1(\omega_n{}a_0)-\frac{J_1(\omega_n)}{Y_1(\omega_n)} Y_1(\omega_n{}a_0)\right].
\end{equation}
For early times, $t\ll{}1$ and therefore
\begin{equation}
    \exp(-\omega_n^2t) \sim 1-\omega_n^2t+\mathcal{O}(t^2),
\end{equation}
so that
\begin{multline}\label{anasolconstrained_supp}
   u_s(t,a_0) \sim \sum_{n=1}^{\infty} t\,f_n\,\left[J_1(\omega_n{}a_0)-\frac{J_1(\omega_n)}{Y_1(\omega_n)} Y_1(\omega_n{}a_0)\right] = t \sum_{n=1}^{\infty} f_n \mathcal{R}_n(a_0) = t\,\left(\frac{q}{a_0}\right).
\end{multline}
The nonlinear models lead to the same prediction for $t\ll{}1$ and $u_s\ll{}1$ since they are all asymptotically equivalent to the fully linear model for small strains. Letting $t=\epsilon{}T$ and $u=\epsilon{}U$ for some functions $T$ and $U$ that are strictly order one, Equations (3.1--3.2) subject to Equations (2.16) and (2.14) reduce at leading order to
\begin{equation}\label{early time}
    \frac{\partial{U}}{\partial{T}}=\frac{q}{r}.
\end{equation}
At $r=a=a_0+u_s(a,t)=a_0+\epsilon{}U$, Equation~(\ref{early time}) becomes at leading order
\begin{equation}\label{early time a}
    \frac{\partial{U}}{\partial{T}}\bigg|_{a}=\frac{q}{a_0} \quad \implies \quad U(a,T)=T\,\left(\frac{q}{a_0}\right) \quad\implies\quad u_s(a,t)=t\,\left(\frac{q}{a_0}\right),
\end{equation}
where we have applied the initial condition that $u_s(a,0)=0$. Hence, the early time behaviour for all models is well approximated by $u(t,a)\sim{}(q/a_0)t$.

\section{Evolution time relative to the L-$k_0$ model}
\setcounter{figure}{0}

\begin{figure}[h]
    \centering
    \includegraphics[width=5in]{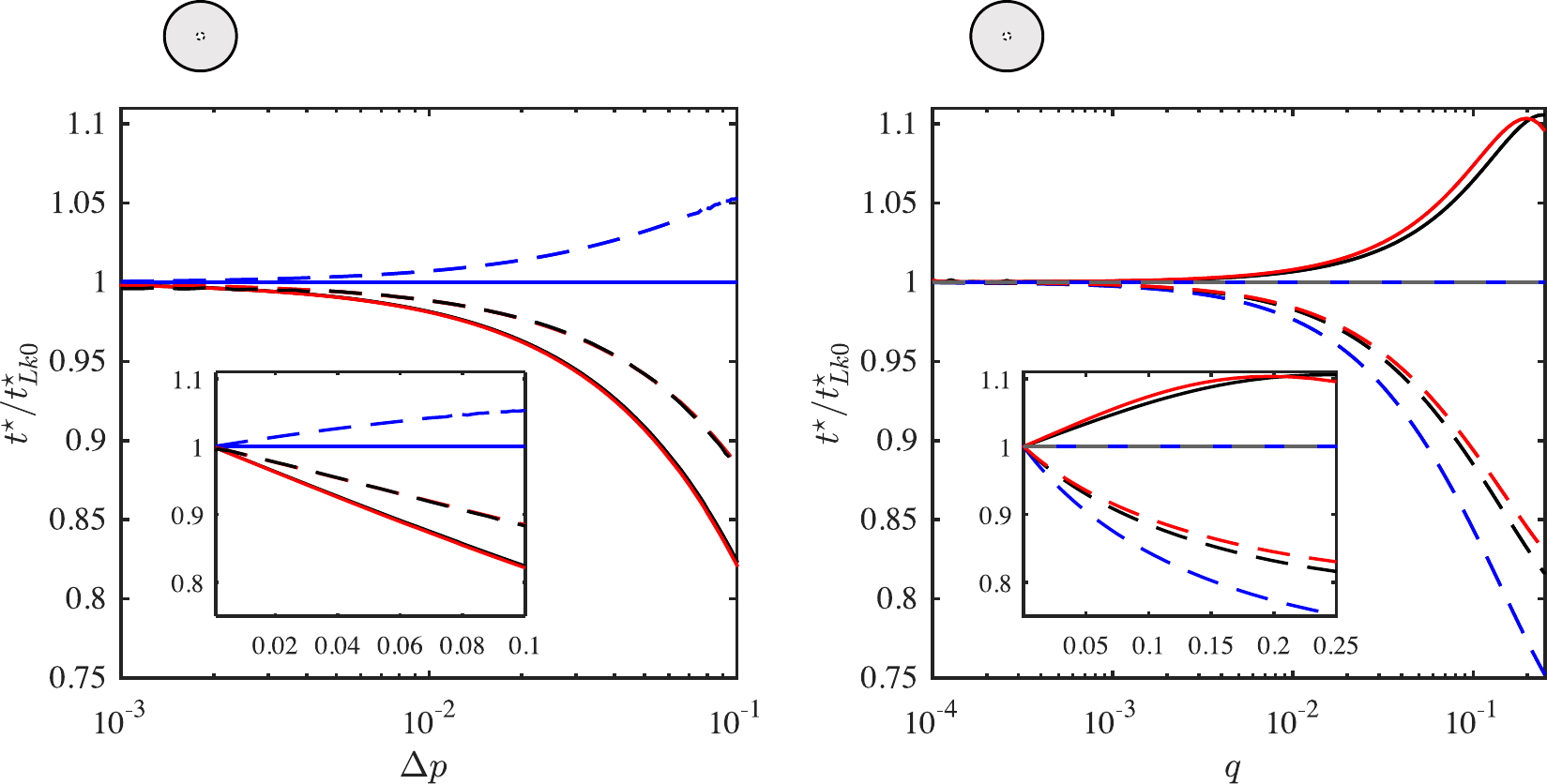}
    \caption{As Figure~5 of the main text, but now normalising the evolution timescale by that of the L-$k_0$ model. \label{qvsdp_tch_supp} }
\end{figure}

\begin{figure}
    \centering
    \includegraphics[width=5in]{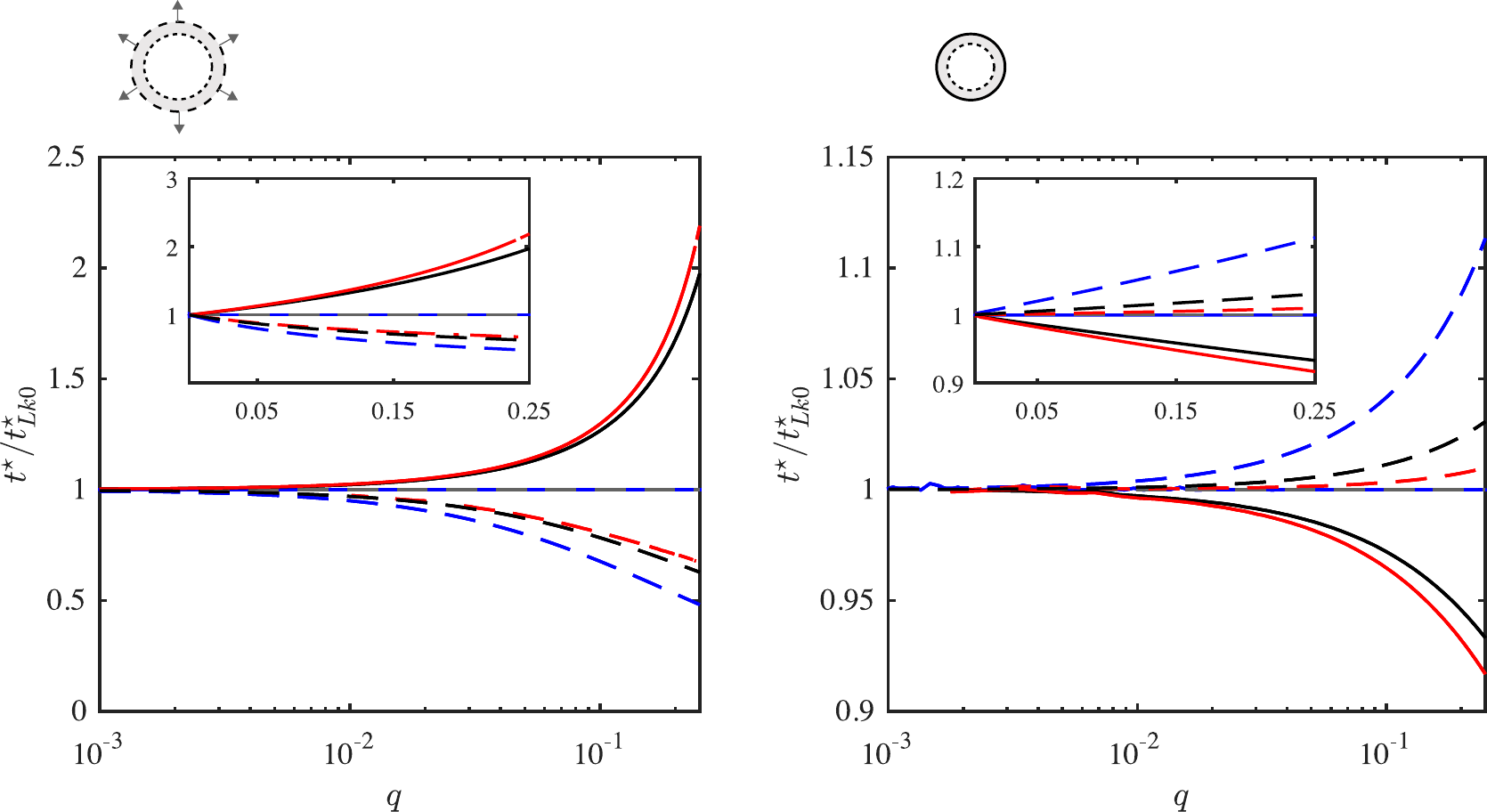}
    \caption{As Figure~8 of the main text, but now normalising the evolution timescale by that of the L-$k_0$ model. \label{ap_vs_zd_tch_supp} }
\end{figure}

\begin{figure}
    \centering
    \includegraphics[width=5in]{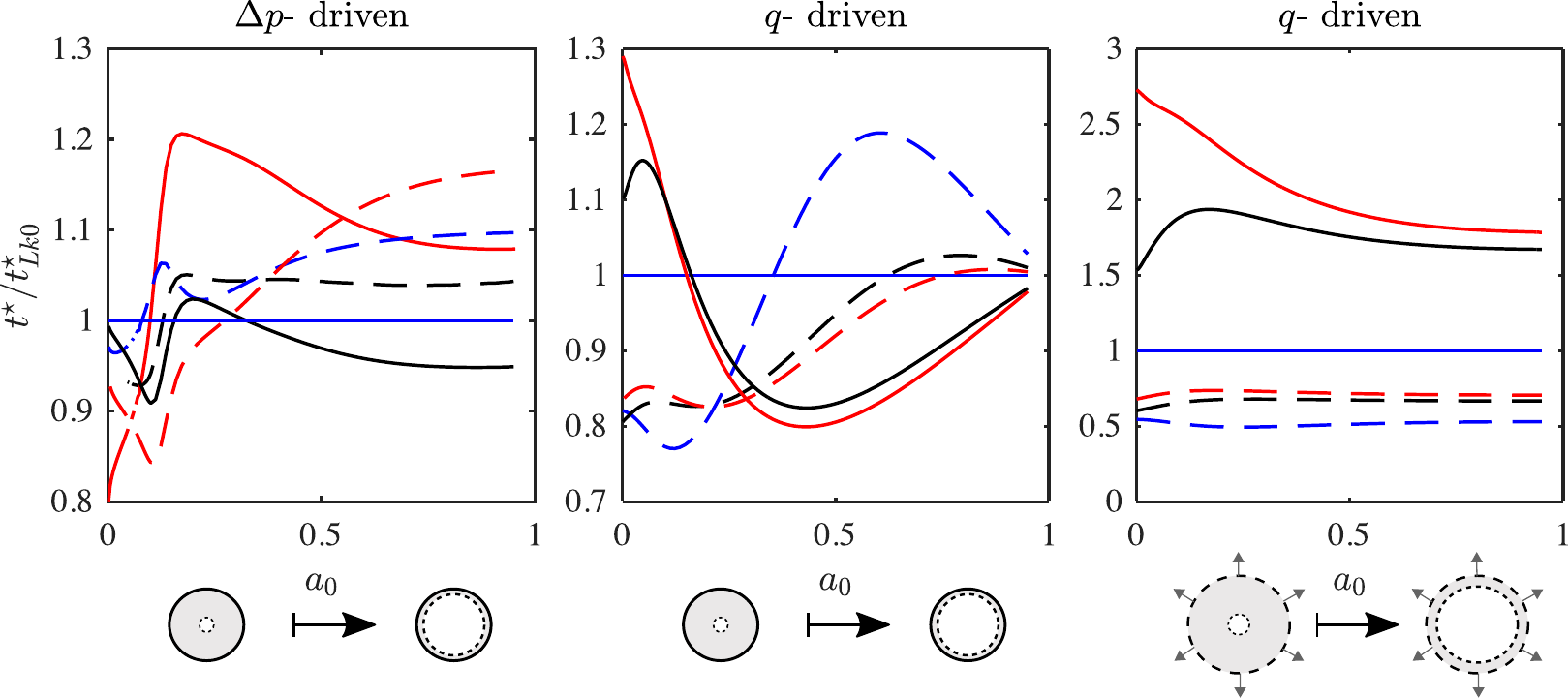}
    \caption{As Figure~9 of the main text, but now normalising the evolution timescale by that of the L-$k_0$ model. \label{t_ch_supp} }
\end{figure}

In Figures~\ref{qvsdp_tch_supp}, \ref{ap_vs_zd_tch_supp}, and \ref{t_ch_supp}, we present the characteristic evolution timescale $t^\star$ as in Figures 5, 8, and 9 of the main text, respectively, but now normalising by the value for the L-$k_0$ model to highlight the degree to which nonlinearity accelerates or decelerates the evolution relative to the L-$k_0$ model.

\clearpage

\section{Impact of constraint on a thick-walled cylinder for fixed $q$}\label{thickwalled}

\setcounter{figure}{0}

\begin{figure}[h]
    \centering
    \includegraphics[width=5in]{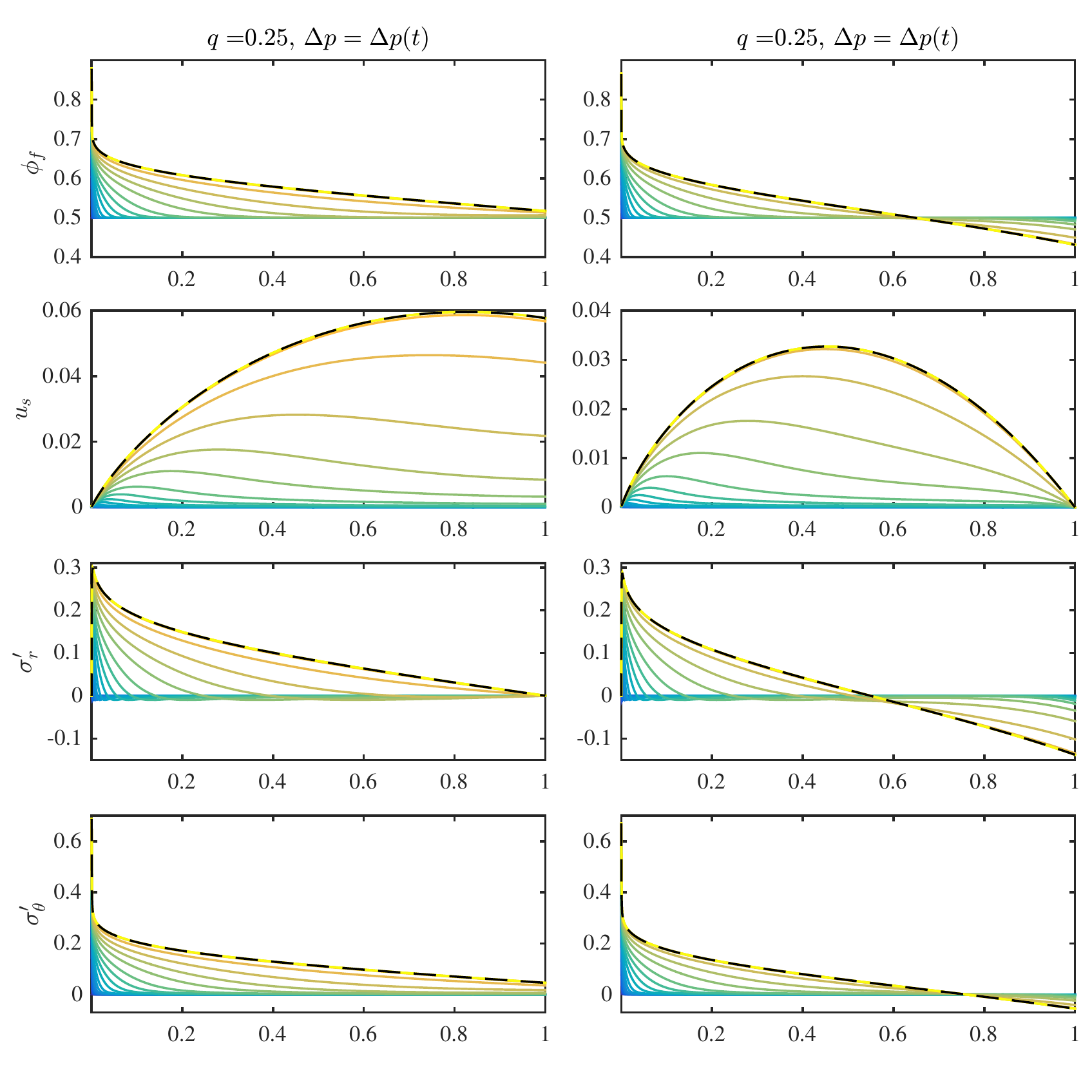}
    \caption{The time evolution of an unconstrained thick-walled cylinder (left column) and a constrained thick-walled cylinder (right column) for the Q-$k_\mathrm{KC}$ model. Both cylinders have $a_0 = 10^{-4}$ and are driven by fixed $q=0.25$. We show the solutions at times logarithmically spaced from $t= 10^{-7}$ (blue) to $t= 10$ (yellow). For reference, we also show the steady state (black dashed lines). The evolution is qualitatively very similar, as are the steady states.}
    \label{all_quants_apvszd2}
\end{figure}

In Figure~\ref{all_quants_apvszd2}, we show the time evolution of an unconstrained thick-walled cylinder (left column) and a constrained thick-walled cylinder (right column), both for $a_0 = 10^{-4}$ and driven by fixed $q=0.25$. The evolution in these two cases is very similar, as are the steady states~[3].


\end{document}